%% file: paper.tex
\documentclass[superscriptaddress,floatfix,nofootinbib,preprintnumbers]{revtex4}

\usepackage{amsmath,graphicx,tabularx,url,times}

\begin{document}

\include{declare}
\title{Measuring Hidden Higgs and Strongly-Interacting Higgs Scenarios}

\author{Sebastian Bock}
\affiliation{Institut f\"ur Theoretische Physik, 
             Universit\"at Heidelberg, Germany}

\author{Remi Lafaye}
\affiliation{LAPP, Universi\'e de Savoie,
             IN2P3/CNRS, Annecy}

\author{Tilman Plehn}
\affiliation{Institut f\"ur Theoretische Physik, 
             Universit\"at Heidelberg, Germany}

\author{Michael Rauch}
\affiliation{Institut f\"ur Theoretische Physik, 
             Karlsruhe Institute of Technology, Karlsruhe, Germany}

\author{Dirk Zerwas}
\affiliation{LAL, IN2P3/CNRS, Orsay, France}

\author{Peter M. Zerwas}
\affiliation{Deutsches Elektronen-Synchrotron DESY, Hamburg, Germany}
\affiliation{Institut f\"ur Theoretische Teilchenphysik und 
             Kosmologie, RWTH Aachen University, Germany} 

\date{\today}


\begin{abstract}
\noindent
{\it Higgs couplings can be affected by physics beyond 
  the Standard Model. We study modifications
  through interactions with a hidden sector and in specific composite 
  Higgs models accessible at the LHC.  Both scenarios give rise to congruent patterns of 
  universal, or partially universal, shifts. In addition, Higgs decays 
  to the hidden sector may lead to invisible decay modes which we 
  also exploit. Experimental bounds on such potential modifications will 
  measure the concordance of an observed Higgs boson with the Standard Model.}
\end{abstract}

\maketitle

\section{Introduction}
\label{sec:intro}

Embedding the Standard Model [SM] into an extended theory will, in
general, modify the couplings predicted in the minimal Higgs sector of
the Standard Model~\cite{higgs_review1,higgs_review2}. Thus, measuring
the Higgs couplings at the LHC~\cite{duehrssen,sfitter_higgs} will
shed light on potential scenarios beyond the Standard Model. In two
well motivated models the analysis is particularly transparent:

First, theories beyond the Standard Model may include a hidden
sector. The Higgs field offers an attractive candidate 
for opening the portal to such a hidden sector~\cite{higgs_portal,wells,bij,bsm_review}. The coupling between 
the SM-singlet Higgs mass term and the corresponding SM-neutral 
Higgs term in the hidden sector leads to an interaction which transfers
the renormalizability from the Standard Model to the extended theory. 

Second, interpreting the Higgs particle as composite pseudo-Goldstone
boson generated by new strong interactions is clearly a
well motivated scenario~\cite{georgi,silh,silh_pheno,MMMM}. The strong interactions 
will induce deviations from the properties predicted for a minimal point-like
Higgs particle so that the Higgs profile may signal dynamical
structures beyond the Standard Model. 
\medskip

If the Standard Model Higgs field couples to a hidden Higgs sector, the couplings
to the other Standard Model particles are modified universally; in addition,
decays into the hidden sector may generate an
invisible decay mode and thereby affect the total width. 
In strong interaction models where the Higgs boson emerges 
as a pseudo-Goldstone boson all the Higgs couplings, cross sections
and partial widths, may be altered universally or following 
a simple fermion {\it vs} boson pattern. Moreover, this scenario
does not predict any novel invisible
decays. The set of these characteristics 
will allow us to discriminate between the two scenarios. 

In general, depending on the
operator basis
chosen~\cite{dieter_op,wbf_op,spin_op,gravitons_op,new_op}, some
$\mathcal{O}(10)$ free parameters may affect the measured production
and decay rates at the LHC. 
A universal [or partially universal] modification of the Higgs couplings tremendously simplifies
the complexity of any experimental analysis to the
measurement of just one, or two, new parameters. Furthermore, setting bounds on universal
deviations from the Standard Model Higgs couplings measures the degree of
concordance between the observed Higgs boson and the Standard Model in
a particularly transparent form. 
\medskip

In the two scenarios introduced above, the twin width-ratios 
of the Higgs boson are modified by a parameter~$\kappa$:
\begin{equation}
\frac{\Gamma_p \Gamma_d}{\gtot} = \kappa \, 
     \left(\frac{\Gamma_p \Gamma_d}{\gtot} \right)^\text{SM}   \,.
\label{eq:ratio}
\end{equation}
The partial widths refer to the production channel $p$ and the decay
mode $d$, either exclusively or summing over sets of initial or final states.
These ratios are measured, at the Born level, directly 
by the product of production cross section times decay branching ratio 
of the process $p \to Higgs \to d$ in the narrow width approximation. 
In the hidden sector the parameter $\kappa$ is universal; in the
strong interaction scenario we consider it
may 
take different values for Higgs couplings to vector bosons or
fermions~\cite{MMMM}. 

For a hidden sector the decay label $d$ includes invisible Higgs decays, 
\ie the partial width $\Gamma_\text{hid}$. This second parameter 
can be measured via the invisible branching ratio $\brinv$.
It is well-known~\cite{inv_higgs,ATLAS-TDR,CMS-TDR,DeRo} that the
determination of $\brinv$ at hadron colliders is quite demanding,
even through it naturally appears in many extensions of the Standard 
Model, like four lepton generations or supersymmetry~\cite{bsm_review}.

In the present study we will show, adopting the tools of SFitter, at which level
$\kappa$ as well as $\Gamma_\text{hid}$, if present, can be determined
at the LHC.
\medskip

The measurements of $\kappa$ and ${\text{BR}}_{\text{inv}}$ do not require
the estimate of the total width appearing in the 
denominator of Eq.(\ref{eq:ratio}). Nevertheless, estimating 
$\Gamma_{\text{tot}}$ will provide us with consistency checks on our 
theoretical ansatz.
One way is to simply identify the total Higgs width with the sum of all 
partial widths, with or without invisible channels~\cite{sfitter_higgs}.
The non-observed partial widths are fixed to the Standard Model value
scaled by the same global factor applied to the observed partial widths. 
This method relies strongly on the recent 
resurrection of the $H \to b\bar{b}$ channel based on fat jet searches~\cite{fat_jets_th,fat_jets_exp}. 
An alternative way to construct an upper limit 
to the Higgs width --- to be combined with the lower limit 
from all observed partial widths --- would be motivated by the unitarization 
of $WW \to WW$ scattering. The Standard Model Higgs state saturates 
this unitarization, so modulo quantum corrections the relation 
$g_{WWH} \lesssim g_{WWH}^\text{SM}$ becomes an upper bound 
to the Higgs width~\cite{duehrssen}. We cannot use such an additional 
constraint because the observed scalar state in our models overlaps 
only partly with the state related to electroweak symmetry breaking.
\medskip

Extracting Higgs parameters from LHC
data~\cite{duehrssen,sfitter_higgs} forces us to pay attention to the
different uncertainties affecting the rate measurements and their
comparison to theory predictions for Higgs
production~\cite{hprod_gg,hprod_wbf} and
decay~\cite{DSZ,Hdecay}.  For typical luminosities around 30~fb$^{-1}$
statistical uncertainties will be the limiting factor for example in
weak-boson-fusion or Higgs-strahlung channels.  Simulating these
statistical uncertainties we use Poisson statistics. Experimental
systematic errors, as long as they are related to measured properties
of the detector, are expected to be dominantly Gaussian. We include
flat theory errors based on the Rfit profile-likelihood
construction~\cite{rfit,sfitter}.

In part of our studies ratios of Higgs couplings will play a crucial
role. Higher precision in measuring these ratios may naively be expected
compared with individual measurements of couplings~\cite{duehrssen}. 
For such an improvement the analysis should not be statistics
dominated, which it largely is however for an integrated luminosity of
30~fb$^{-1}$. Moreover, while experimental systematic uncertainties tend to cancel
between the same Higgs decays but different production channels, the dominant theory
errors are expected to cancel for identical production mechanisms.  In
line with these arguments we have found that using ratios does not
significantly improve the results of Higgs sector
analyses~\cite{sfitter_higgs}.
\medskip

In
this study we will show how
$\kappa$ as well as $\Gamma_\text{hid}$ can be determined 
using SFitter.
Starting from the completely exclusive likelihood map, SFitter
determines the best-fitting point in the Higgs-sector parameter
space. While a Bayesian probability analysis of the entire Higgs
parameter space at the LHC is spoiled by noise, profile
likelihoods can be studied in the vicinity
of the best-fitting points~\cite{sfitter_higgs}.  In this analysis we
assume that we already know the global structure of the likelihood
map, so we can focus on the local properties around the SM-like
solution.  As it will turn out, alternative solutions can be studied
nevertheless, for example with sign switches for some
of the Higgs couplings.

Technically, the analysis presented in this paper is based on the
SFitter-Higgs setup. The Higgs production rates include NLO QCD
corrections except for the top-quark associated production mode. For the
decays we use a modified version of HDECAY~\cite{Hdecay}, which contains
both NLO QCD terms and off-shell decays into vector bosons.  For a list
of all measurements
and their different errors we refer to Ref.~\cite{sfitter_higgs}. 
Compared to this previous analysis we have updated the numbers for the
$H \to b\bar{b}$ channel in associated production with vector bosons
from the recent ATLAS study~\cite{fat_jets_exp}, which confirms the previously
obtained significances. The event rates for weak-boson-fusion production
with decay into invisible states are adopted from Ref.~\cite{ATLAS-TDR}.
The central data set is smeared around the theory predictions 
according to the theoretical error and the experimental errors, taking into account the 
correlations among the observables. 
For each of the toy-experiments we determine the best-fit values.  This numerical determination
of the resulting parameter uncertainties 
is fitted to Gaussian distributions. 

The new technical aspect of the present study is the more refined approach to
the hypotheses tested: if we do not measure all Higgs couplings
independently but instead test a given model hypothesis, the limits on
the extracted model parameters improve significantly. Because this
approach requires fewer measurements we now consider Higgs masses
between 110 and 200~GeV and find a significant enhancement of the
determination power for 30~fb$^{-1}$ of LHC data at a collider energy of
14~TeV.
\medskip

\section{Higgs Portal to Hidden Sector}
\label{sec:portal}

The Standard Model, or extensions of it, may be connected to a hidden
sector.  An interesting realization of such a mechanism is provided by
specifying the scalar Higgs domains in both sectors as the link between 
the two sectors~\cite{higgs_portal,wells}. To
explore the possibility of detecting a hidden sector at the LHC we
investigate a scenario in which the Standard Model Higgs sector is
coupled to the hidden Higgs sector through quartic interactions. Such a
scalar system is technically transparent and may therefore serve as
paradigm for generic experimental features that could signal a hidden
sector. There are many variants to this specific scenario, {\it e.g.} a
hidden scalar sector without spontaneous symmetry breaking, large
ensembles of scalar fields, etc. The scenarios can be disentangled by
analyzing a few characteristic observables of the Higgs particles, in
particular Higgs couplings. In this letter we will concentrate on the
simplest setup to quantify the potential of experimental analyses at the
LHC.
\medskip

The scenario we will focus on for now is described by the Higgs
potential of the Standard Model $[s]$, the isomorphic potential in the
hidden sector $[h]$, and the quartic interaction potential coupling 
the two sectors with strength $\eta_\chi$, {\it videlicet},
\begin{equation}
 \mathcal{V} =
 \mu^2_s |\phi_s|^2 + \lambda_s |\phi_s|^4 
 \; + \;
 \mu^2_h |\phi_h|^2 + \lambda_h |\phi_h|^4
 \; + \;
 \eta_\chi |\phi_s|^2 |\phi_h|^2 \, .
\end{equation}
Expanding the two Higgs fields about their vacuum expectation values
$\phi_j \rightarrow (v_j + H_j)/\sqrt{2}$ we encounter shifts away 
from the standard values by the interaction term,
\begin{equation}
v_s^2 = 
\frac{1}{\lambda_s} \,
\left(- \mu_s^2 - \dfrac{1}{2}\eta_\chi v_h^2 \right)
\qquad \text{and} \qquad 
v_h^2 = 
\frac{1}{\lambda_h} \,
\left(- \mu_h^2 - \dfrac{1}{2}\eta_\chi v_s^2 \right) \; .
\end{equation}     
The Higgs states in the SM and the hidden sector will be mixed. Diagonalizing the
Higgs mass matrix,
\begin{equation}
   {\mathcal{M}}^2 = \left( \begin{array}{cc}
                               2 \lambda_s v^2_s    &   \eta_\chi v_s v_h    \\   
                               \eta_\chi v_s v_h    &   2 \lambda_h v^2_h
                             \end{array} \right)                             \,,
\end{equation}
generates two mass eigenvalues $M_{1,2}$ and the mixing angle $\chi$
\begin{eqnarray}
   M^2_{1,2} &=&   (\lambda_s v^2_s + \lambda_h v^2_h)
                    \pm \left[ (\lambda_s v^2_s - \lambda_h v^2_h)^2 
                    + (\eta_\chi v_s v_h)^2 \right]^{1/2}              \nonumber \\
\tan \, 2\chi &=& \eta_\chi v_s v_h / [\lambda_s
                v^2_s - \lambda_h v^2_h]
\end{eqnarray}
for the mass eigenstates
\begin{alignat}{5}
   H_1 &=&  \cos\chi \, H_s + \sin\chi \, H_h &    \notag \\
   H_2 &=& -\sin\chi \, H_s + \cos\chi \, H_h & \; .
\end{alignat}
Both, $H_1$ and $H_2$ couple to Standard Model fields through their $H_s$ 
components and
to the hidden sector through the $H_h$ admixtures.  To focus on generic
features  we assume the potential
parameters $\lambda_j$ and $v_j$ to be of similar size and the mixing
parameter $\eta_\chi$ to be moderate. The properties of $H_1$ then remain
dominated by the Standard Model component, while the properties of $H_2$
are characterized primarily by the hidden Higgs component.\medskip

The phenomenology of a Higgs portal to the hidden sector depends
on whether the standard Higgs particle is lighter or heavier than
the new companion.
In this study we assume that $H_1$ is light and mainly decays 
into Standard Model particles, at a rate reduced by mixing,
and with an admixture of invisible decays to the
hidden sector.  
In general, the heavier $H_2$ bosons decay primarily into particles
of the hidden sector, and only a small fraction by mixing to Standard
Model particles and to light $H_1$ pairs. The production rate of $H_2$,
mediated by mixing, is small. In models in which the $H_2$ decay channels
to the hidden sector are shut, $H_2$ will decay to SM particles with
characteristics quite distinct from $H_1$ decays as the invariant
masses of the decay final states will be different for $H_1$ and $H_2$.
$H_1$ bosons in turn will not decay in such scenarios into novel invisible
channels and decays through fluctuations to virtual $H_2$ states back to
the Standard Model are doubly suppressed. Therefore we will focus on the
light Higgs boson $H_1$ with properties closely related to the Standard
Model{\footnote{Comprehensive analyses of the $H_2$ boson are presently
in progress.}}.

All $H_1$ couplings to Standard Model particles are universally
suppressed by the mixing parameter $\cos\chi$.
In addition,
$H_1$ may decay invisibly 
into the hidden sector. These two features imply 
\begin{alignat}{5}
\sigma           &= \cos^2\chi \, \sigma^\text{SM} \notag \\
\Gamma_\text{vis} &= \cos^2\chi \, \Gamma^\text{SM}_\text{vis} \notag \\ 
\Gamma_\text{inv} &= \cos^2\chi \, \Gamma^\text{SM}_\text{inv} 
                             + \Gamma_\text{hid}  \; .
\label{eq:paras}
\end{alignat}
The two
parameters $\cos \chi$ and $\Gamma_\text{hid}$ will be 
determined in our LHC analysis. $\Gamma_\text{inv}^\text{SM}$ is
generated by Higgs decays $H \to ZZ \to 4\nu$ with an invisible $Z$ branching ratio 
of $4\%$. If invisible Higgs decays will be observed, 
this $4 \nu$ rate can be predicted from observed decays $H \to ZZ 
\to 4\ell$ and can thus be subtracted from the new-physics signal. 
For the sake of simplicity we will omit $\Gamma_\text{inv}^\text{SM}$ from now on.
\medskip

\begin{figure}[b]
\includegraphics[width=0.40\textwidth]{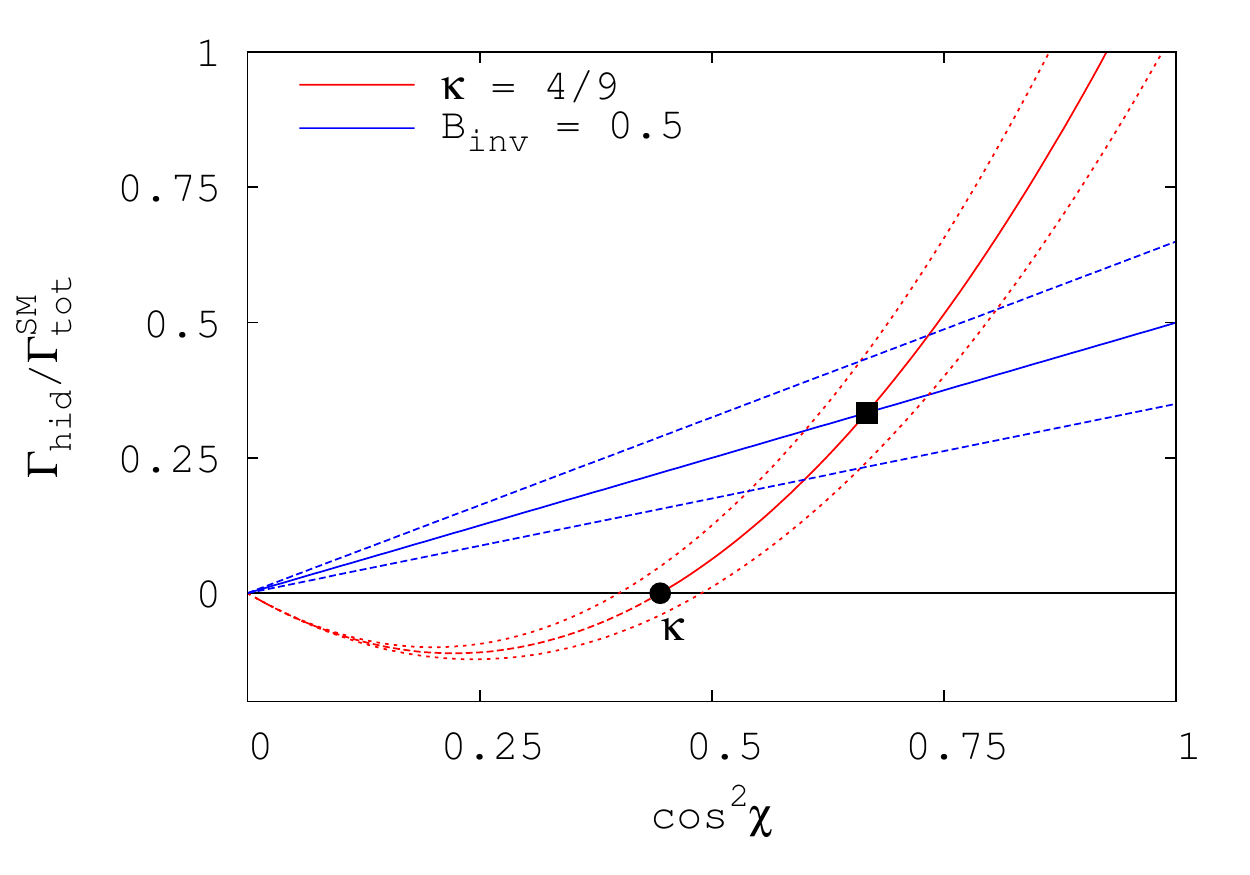}
\vspace*{-5mm}
\caption{Correlations between $\Gamma_\text{hid}$ and $\cos^2\chi$ as
  defined in Eqs.(\ref{eq:paras}), based on measuring $\kappa$ and
  $\bratio$. The two parameters are set to $\kappa = 4/9$ and $\bratio = 0.5$, respectively, for illustration.
  The square marks the final solution of $\cos^2 \chi = 2/3$ and $\Gamma_\text{hid}/\Gamma^\text{SM}_\text{tot} = 1/3$ 
  for this parameter set. The estimated 95\%~CL error bands are explained in the text.}
\label{fig:corr}
\end{figure}

If the invisible decay channel is open, the $\kappa$ parameter 
in the twin ratio Eq.(\ref{eq:ratio}) reads in terms of the
parameters $\cos \chi$ and $\brinv$ [as generated by 
$\Gamma_\text{hid}$]:
\begin{equation}
\kappa = \frac{\cos^2\chi}{1+\bratio} \leq 1       
\quad \text{with} \quad
\bratio  = 
\frac{\brinv}{\text{BR}_{\text{vis}}} =
\dfrac{\brinv}{1 - \brinv}  \,. 
\label{eq:obs}
\end{equation}
$\bratio$ can also 
be expressed as the ratio of invisible decays to 
just one visible decay channel $d$,
$\bratio={\brinv}/({\text{BR}_d/\text{BR}^\text{SM}_d})$. It can thus 
be observed without explicit reference to the total width. 
For $\sin \chi \ll 1$ $\bratio$ 
simply approaches $\brinv$.
\medskip
 
For the Higgs portal the parameters $\kappa$ and $\bratio$ are independent 
of the analysis channels. As a result, ratios of visible branching ratios
are not modified 
${\text{BR}}_{d_1}/{\text{BR}}_{d_2} 
  = (\text{BR}_{d_1}/\text{BR}_{d_2})^\text{SM}$.
The observation of these identities provides a necessary consistency test
for the hidden Higgs scenario. 
\medskip

The detailed experimental analysis of such a scenario will proceed 
in two steps: First, as long as only $\kappa$ is measured but invisible 
Higgs decays are not, upper bounds on the mixing and, in parallel, 
on the fraction of invisible $H_1$ decays can be established,
\begin{equation}
\sin^2\chi \leq 1-\kappa 
\quad \text{and} \quad 
{\text{BR}}_{\text{inv}} \leq 1-\kappa         \,, 
\label{eq:bounds}
\end{equation}
constraining the potential impact of the hidden sector on the
properties of the SM-type Higgs boson. 
With $\kappa$ measured, the hidden $H_1$ partial width will 
be correlated with the mixing parameter
\begin{equation}
\frac{\Gamma_\text{hid}}{\gtot^\text{SM}} =
     \cos^2\chi \, \left( \frac{\cos^2\chi}{\kappa} - 1 \right) \,,
\label{eq:kap}
\end{equation}
as illustrated in Figure~\ref{fig:corr} for $\kappa = 4/9$. For
illustration purposes, the error on $\kappa$ in this figure is chosen 
to be 10\%.

Even though $\bratio$ can be in principle be determined
experimentally, precise measurements of the invisible decay
mode are difficult at the LHC. To show the correlation of the two 
parameters with the two observables we choose 
$\bratio = 0.5$ with a relative error of 30\%, 
as expected for the
final integrated luminosity of 300~fb$^{-1}$~\cite{inv_higgs,ATLAS-TDR,CMS-TDR,DeRo}. Combining Eq.(\ref{eq:kap})
with the definition Eq.(\ref{eq:obs}),
\begin{equation}
\frac{\Gamma_\text{hid}}{\gtot^\text{SM}} = \cos^2\chi \, \bratio \,, 
\end{equation}
allows us to determine both the mixing parameter $\cos\chi$ and the
$H_1$ partial width to the hidden sector $\Gamma_\text{hid}$ individually, {\it cf.}  Figure~\ref{fig:corr}:
\begin{eqnarray}
\cos^2\chi &=& \dfrac{\kappa}{1-\brinv} \qquad {\rm{and}} \qquad     
\dfrac{\Gamma_\text{hid}}{\gtot^\text{SM}} =
      \dfrac{\kappa \, \brinv}{(1-\brinv)^2}             \,.           
\end{eqnarray}
The bands around the intersection
point and their projections on the axes indicate the 95\% CL for the parameters $\cos^2\chi$ and
$\Gamma_\text{hid}$.
\medskip

\begin{figure}[t]
\begin{center}
\includegraphics[width=0.32\textwidth]{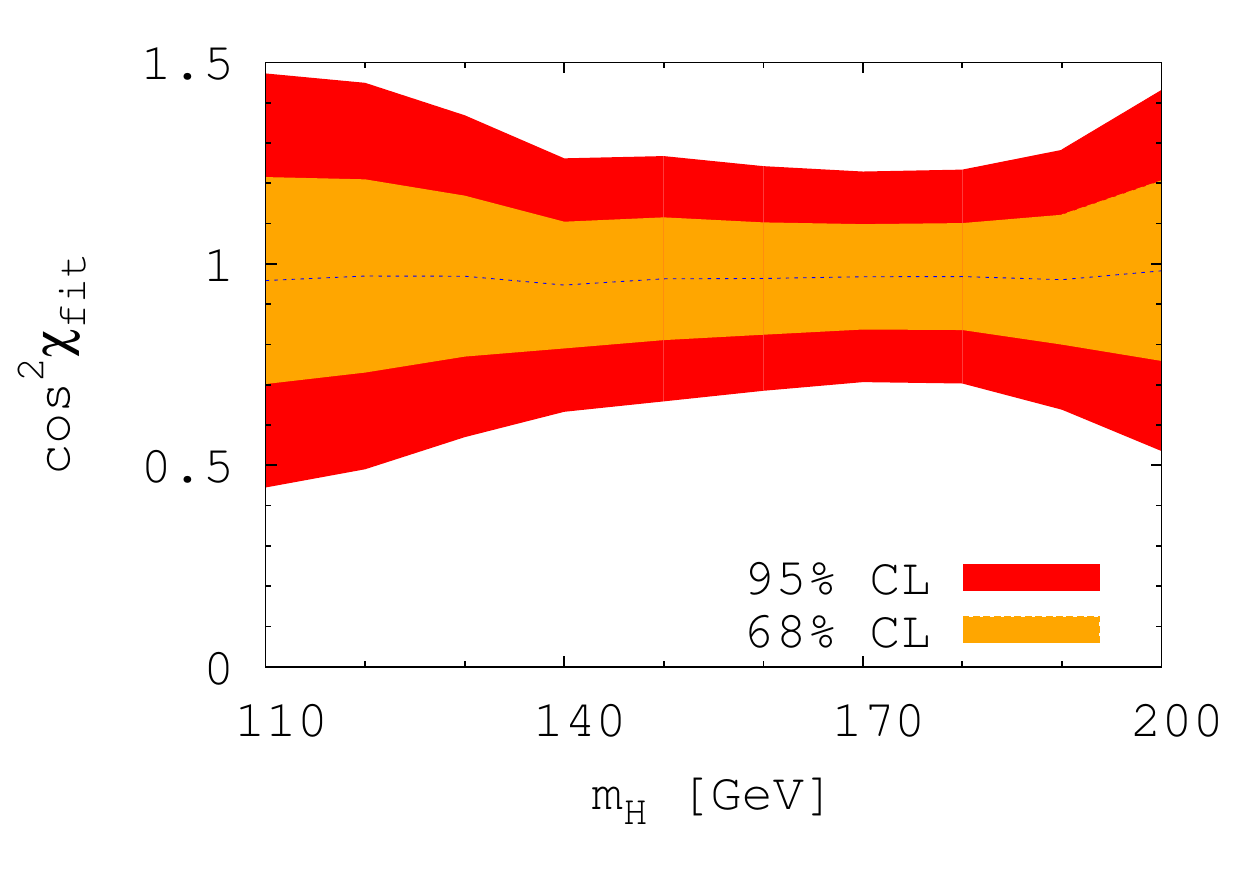}
\includegraphics[width=0.32\textwidth]{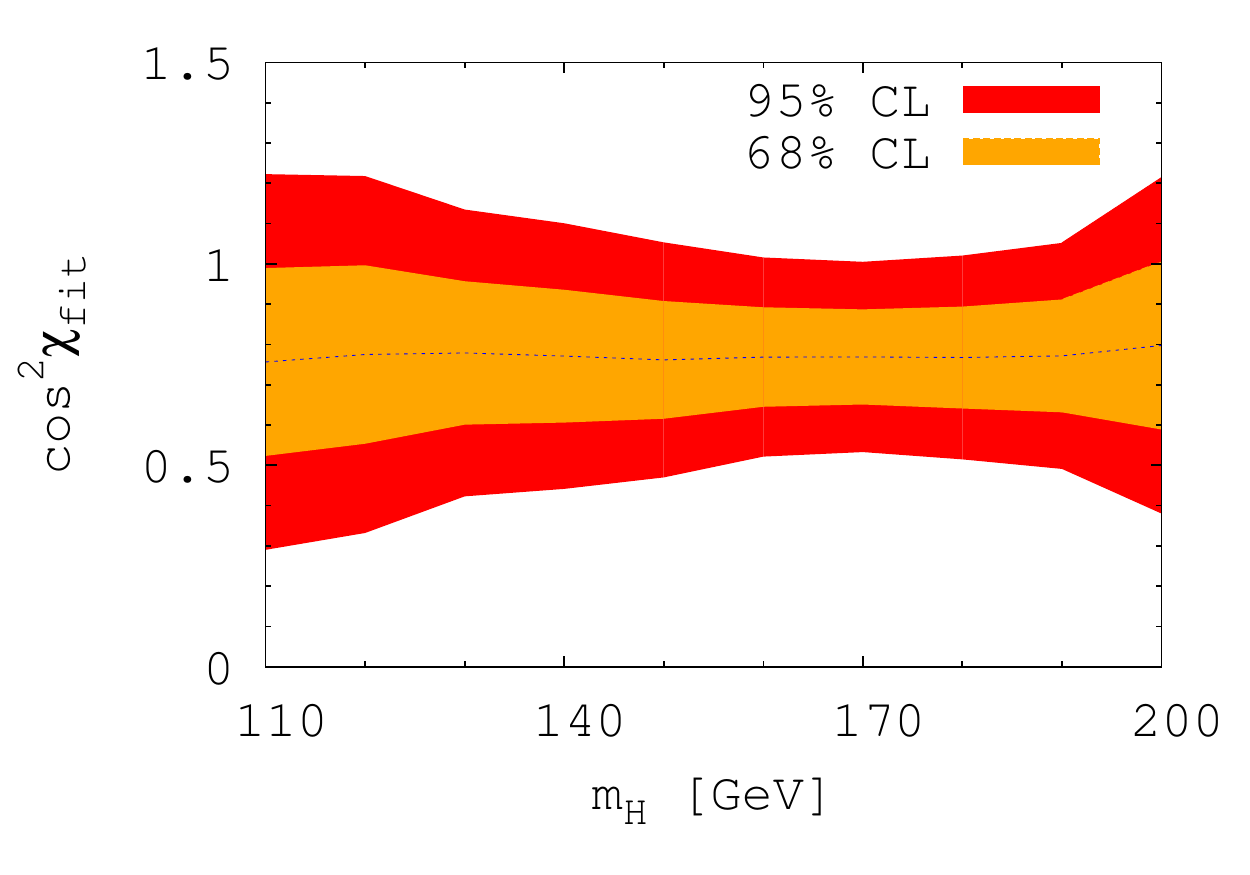}
\includegraphics[width=0.32\textwidth]{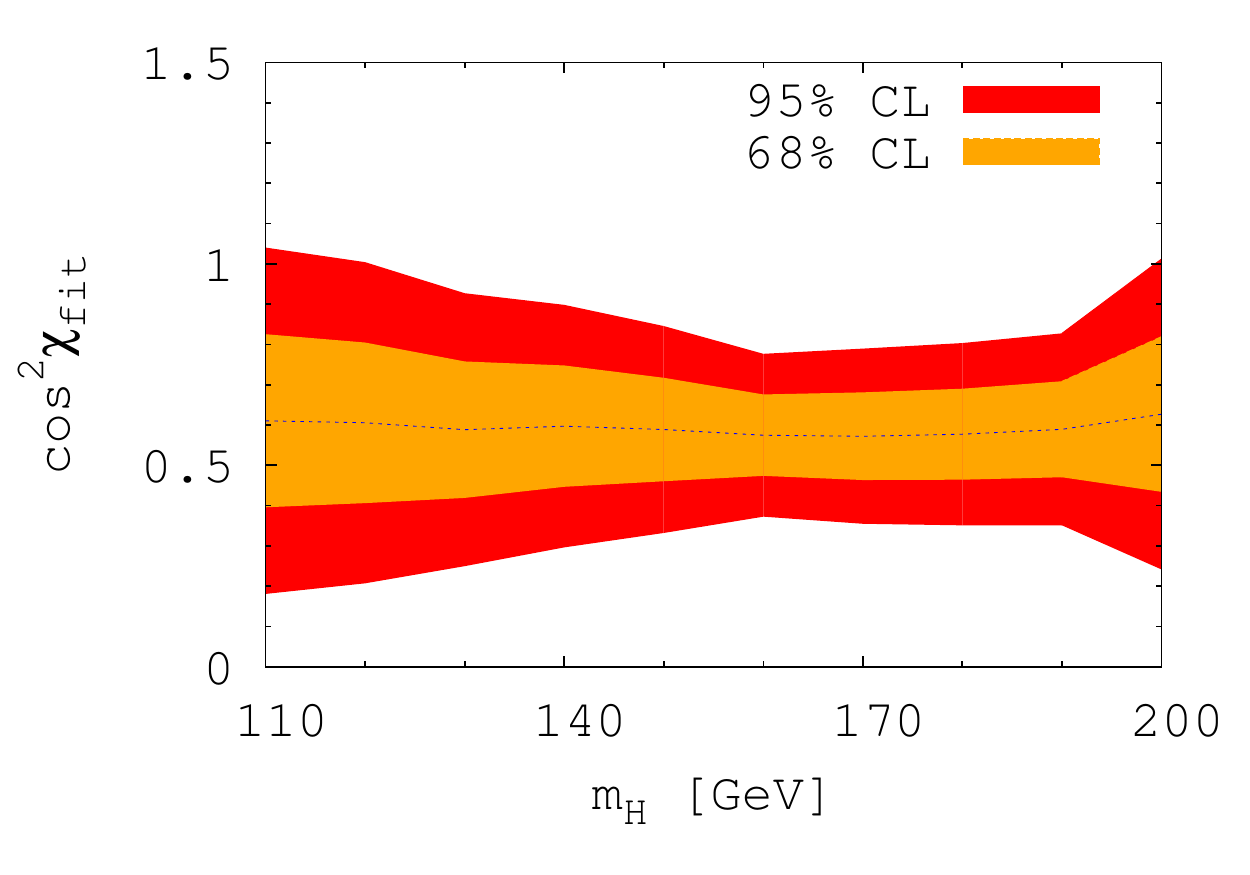} \\[-2mm]
\includegraphics[width=0.32\textwidth]{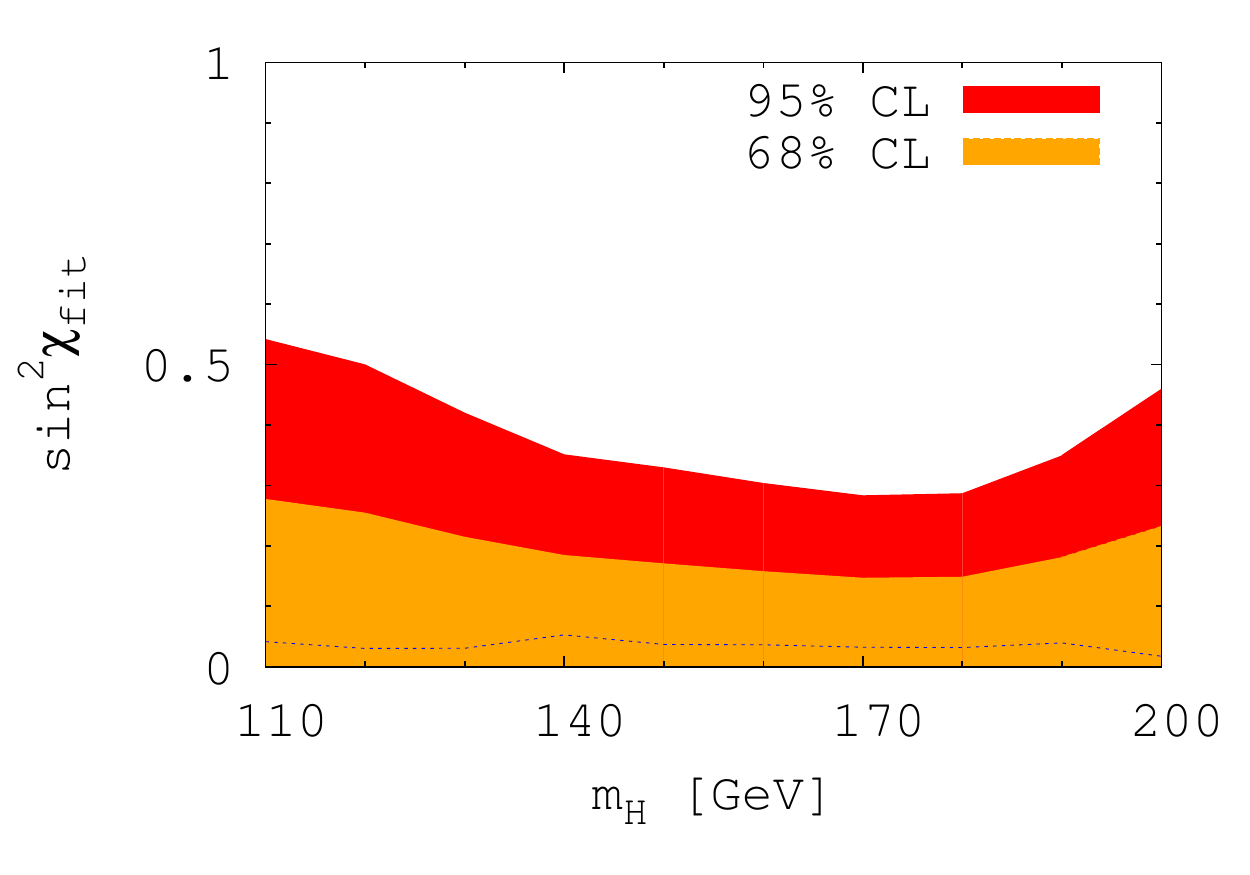} 
\includegraphics[width=0.32\textwidth]{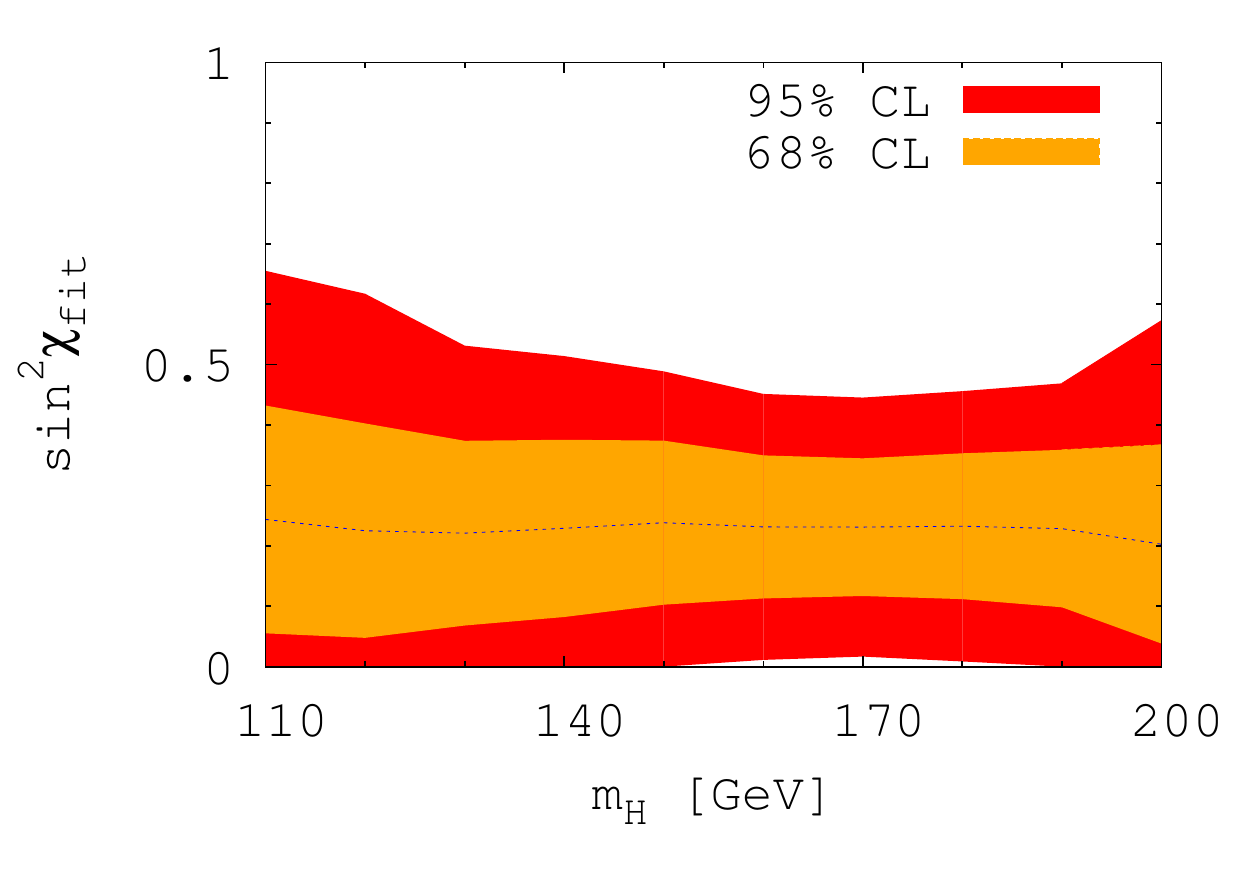}
\includegraphics[width=0.32\textwidth]{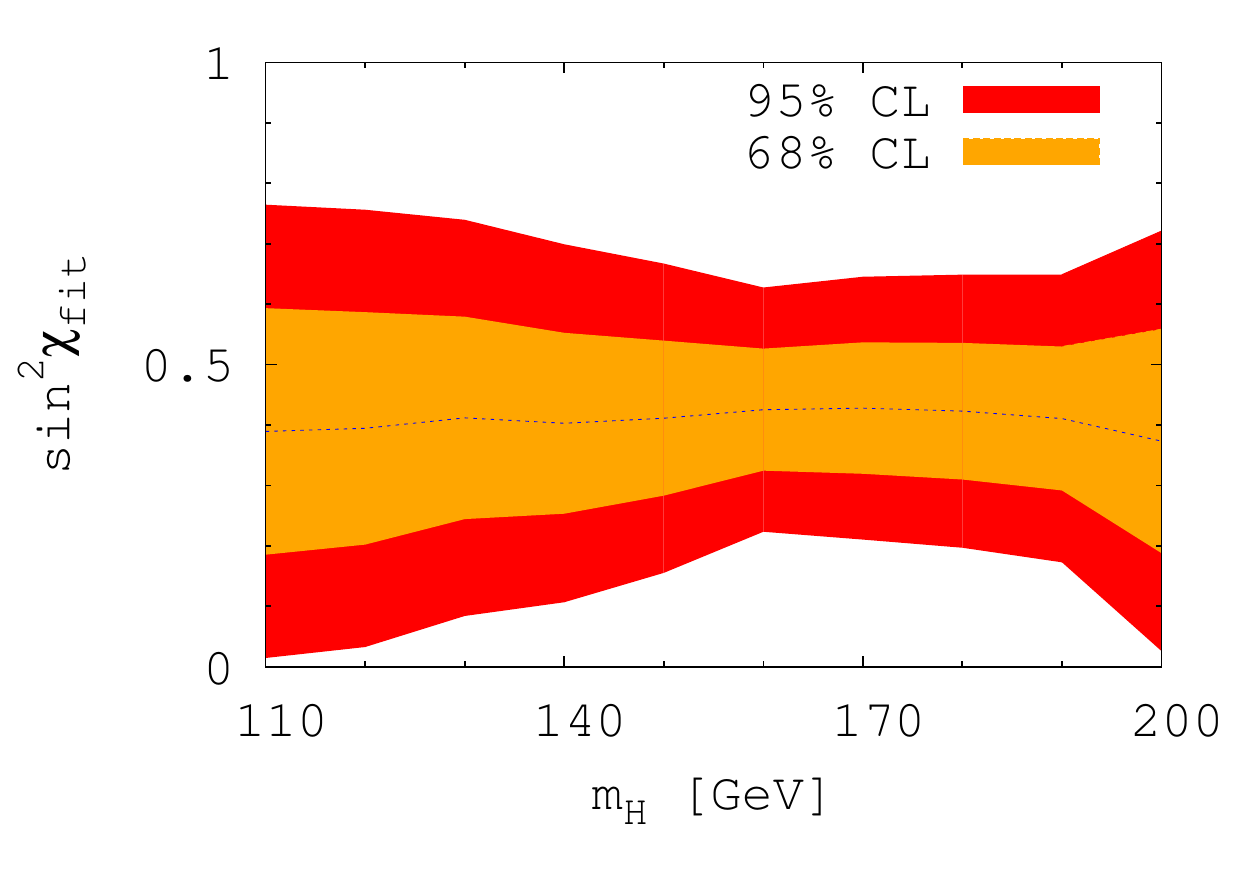} 
\end{center}
\vspace*{-8mm}
\caption{LHC sensitivity to modified Higgs couplings 
  and no invisible decays $\Gamma_\text{hid} = 0$, based on
  30~fb$^{-1}$ of data. Upper: measurement errors as a function 
  of the Higgs mass for $\cos^2   \chi_\text{th}=1.0$ [left], 
  $\cos^2 \chi_\text{th} = 0.8$ [center] and $\cos^2 
  \chi_\text{th} = 0.6$ [right].  Lower: resulting upper 
  and lower bounds 
  on the mixing parameter $\sin^2\chi$, constrained to the 
  physical range.}
\label{fig:portal_noinv}
\end{figure}

\begin{table}[b]
\begin{tabular}{|l|l|l||l|l|l|l|}
\hline                                                                                  
$M_H$ [GeV] & \multicolumn{2}{c||}{ $\;\kappa >\;$ }     & 
              \multicolumn{2}{c|}{ $\;\sin^2\chi, \brinv <\; $ } &
              \multicolumn{2}{c|}{$\;\Gamma_{\text{hid}}/\Gamma^{\text{SM}}_{\text{tot}} <\; $} \\ 
\hline\hline 
 120        & 0.50 & 0.76 & 0.50 & 0.24 & 1.0  & 0.32 \\
 160        & 0.70 & 0.82 & 0.30 & 0.18 & 0.43 & 0.22 \\
 200        & 0.54 & 0.73 & 0.46 & 0.27 & 0.85 & 0.37 \\
\hline
\end{tabular}
\caption{Upper bounds [95\% CL] on mixing and invisible decays expected
  to be set in a Higgs sample within the Standard Model for integrated
  luminosities of 30 [left] and 300 fb$^{-1}$ [right].}
\label{tab:SM}
\end{table}

In a Higgs sector likelihood analysis we can ask three kinds of
questions: 
\begin{enumerate}
\item Can we determine a non-zero mixing $\kappa \neq 1$ from a
  sizable Higgs sample at all?
\item To shut a Higgs portal, what size of $\kappa$ can we exclude if
  we observe Standard Model couplings within given experimental and
  theory errors?
\item To verify the Higgs portal, which finite values of $\kappa$ can
  we establish as a deviation from the Standard Model within
  errors. How do invisible Higgs decays affect this measurement?
\end{enumerate}
\medskip

To answer the first question, the upper panels in Figure~\ref{fig:portal_noinv} show  the
bounds on $\kappa = \cos^2 \chi$ which we can set by analyzing a sample of
Standard Model Higgs bosons with masses between 110 and 200~GeV.
Invisible Higgs decays we neglect in this first step, but all 
standard Higgs 
search channels are exploited based on an integrated luminosity 
of 30~fb$^{-1}$~\cite{sfitter_higgs}.  Starting with the 
Standard Model hypothesis at $m_H = 120$~GeV we can measure $\cos^2 \chi 
= 1 \pm 24\% (48\%)$ at the $68\% (95\%)$~CL. For larger Higgs
masses the error bar improves by a factor of two, due to an increased
statistics of the comparably clean $WW$ and $ZZ$ channels. For even
larger Higgs masses around 200~GeV the over-all event rate drops
again, increasing the measurement error again. As shown in the lower panels of 
Figure~\ref{fig:portal_noinv} we can also translate the 
minimal values of
$\kappa$ into maximal values of the mixing parameter
$\sin^2\chi$ according
to Eq.(\ref{eq:bounds}). In contrast to the upper limits on $\cos^2 \chi$ the
limits on $\sin^2 \chi$ include the constraint $0 \leq \sin^2 \chi \leq 1$.
In Table~{\ref{tab:SM}} we collect the bounds on a modified Higgs
coupling, including the constraint, for three Higgs masses, based on
an integrated luminosity of 30~fb$^{-1}$ as well as expectations for an
integrated luminosity of 300~fb$^{-1}$. Even without observing invisible
decays explicitly, we can translate these results into upper bounds on
the invisible decay width according to Eq.(\ref{eq:obs}). These bounds
are shown in the right column of the table.

The error on the scaled Higgs couplings includes
another square root, which translates into a relative error
$(\Delta g)/g \sim 10\% (20\%)$. These limits measure 
how well the Higgs mechanism in the Standard
Model can be established quantitatively, and they have to be compared 
to an independent variation of all Higgs couplings, which for $m_H=120$~GeV
are expected to be measured to $\mathcal{O} (25\%-50\%)$~\cite{sfitter_higgs}. 
The sizable improvement of the constrained analysis arises first of all because
all experimental channels now contribute to the same measurement, and
secondly because they also determine the total Higgs width much more
precisely than the $H \to b \bar{b}$ fat-jet analysis.
\medskip

\begin{figure}[t]
\begin{center}
\includegraphics[width=0.31\textwidth]{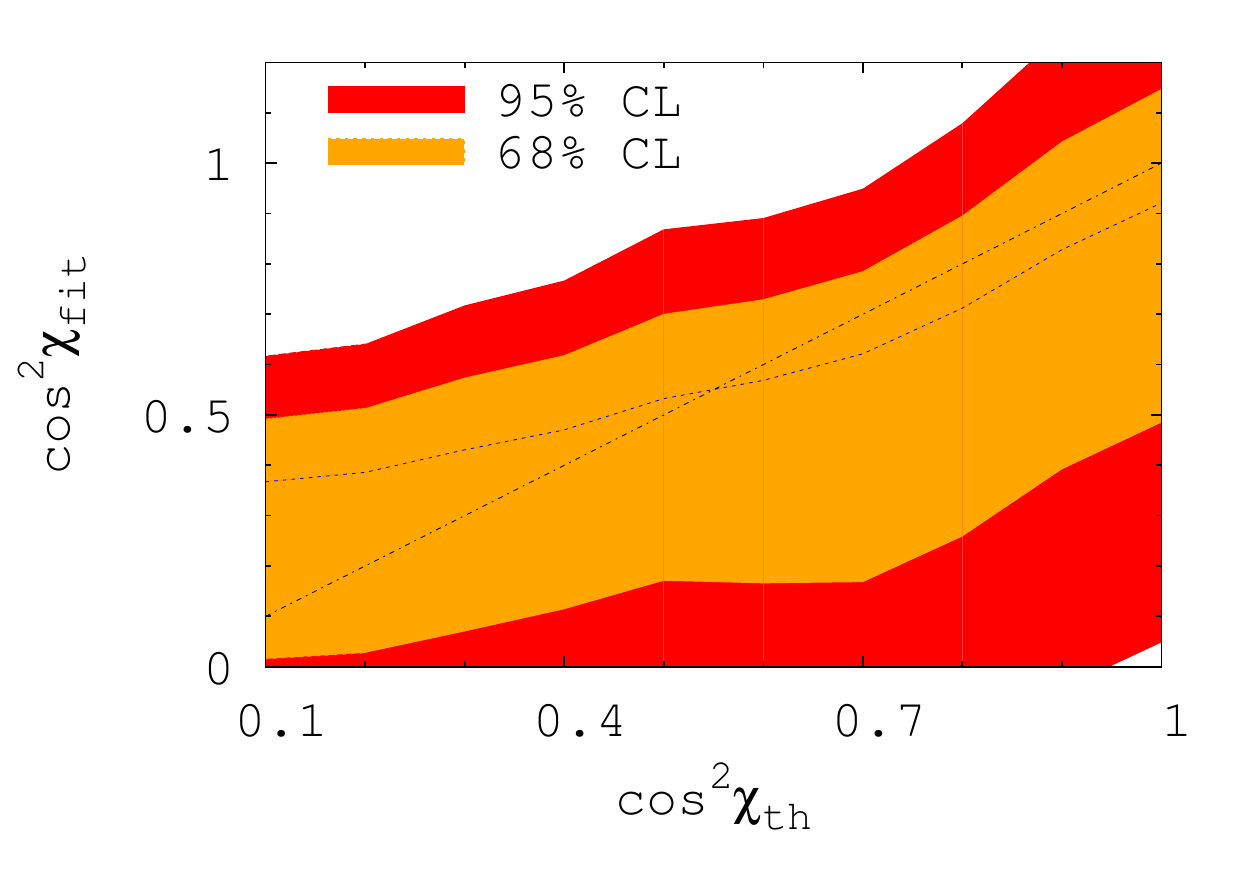}
\includegraphics[width=0.31\textwidth]{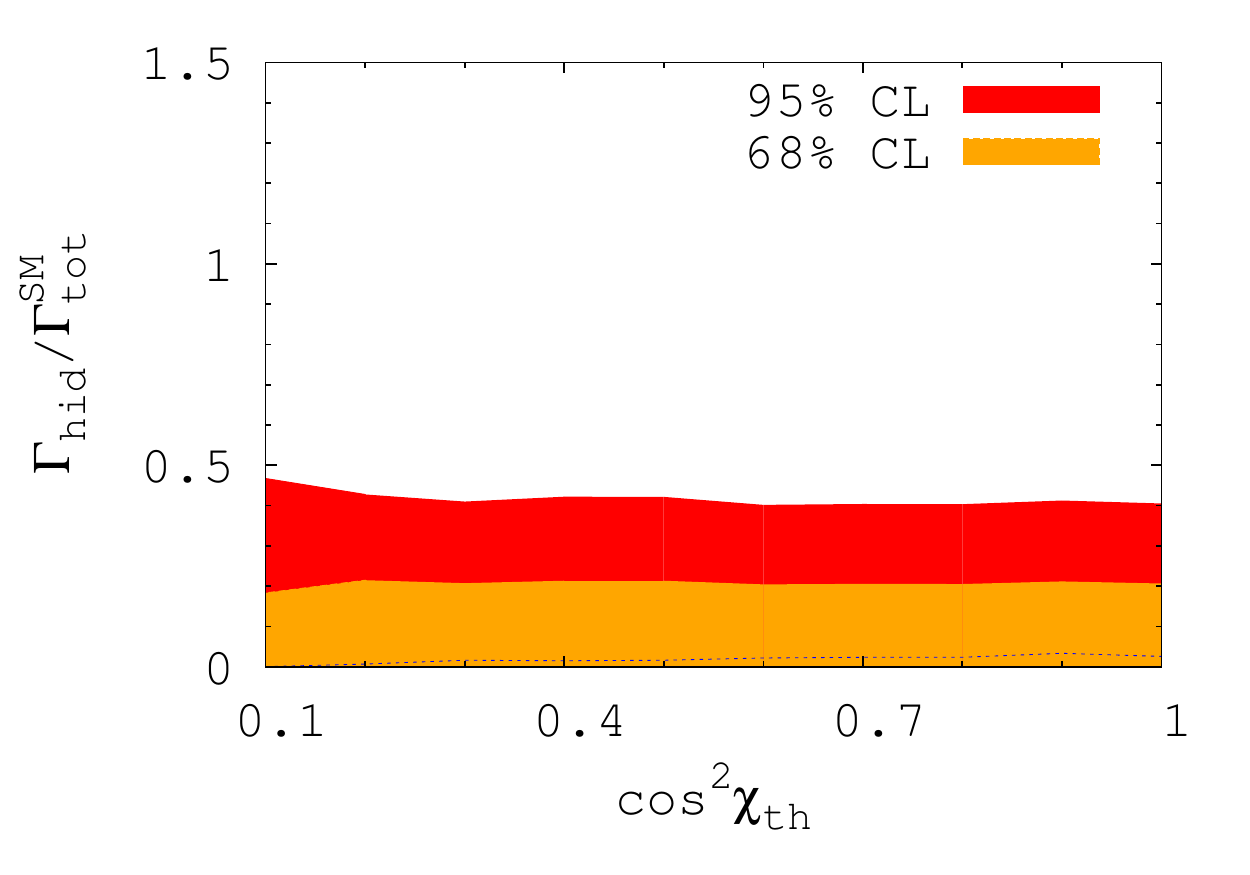} 
\includegraphics[width=0.36\textwidth]{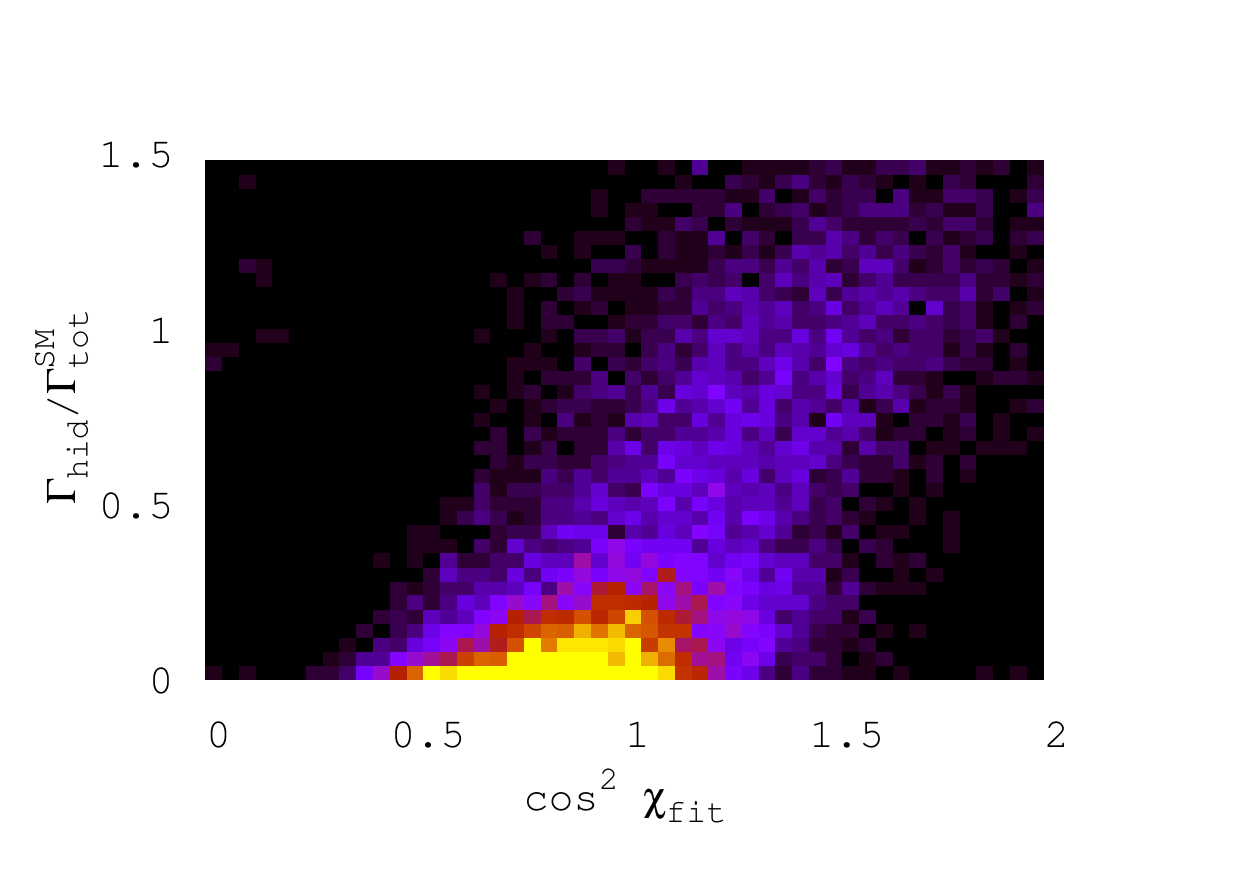} \\[-8mm]
\includegraphics[width=0.31\textwidth]{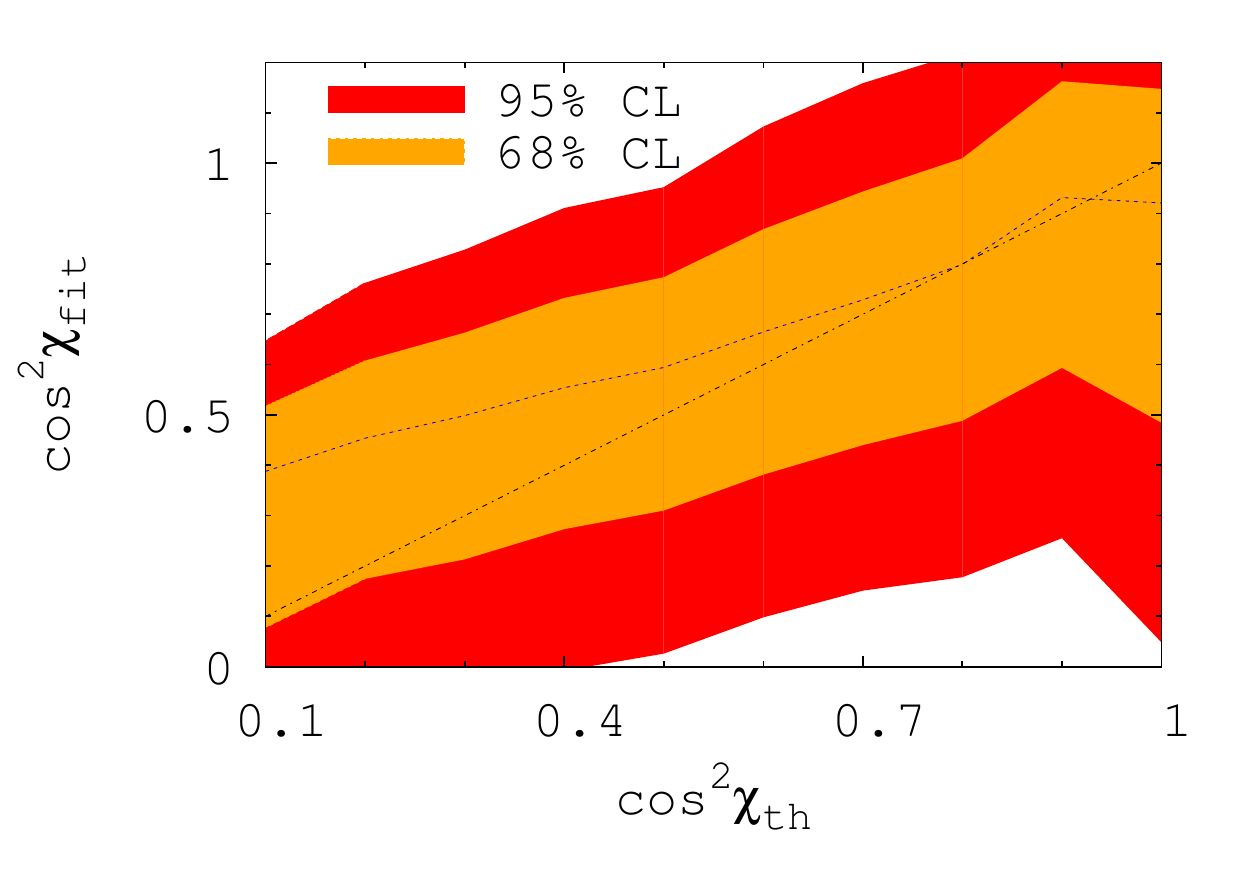} 
\includegraphics[width=0.31\textwidth]{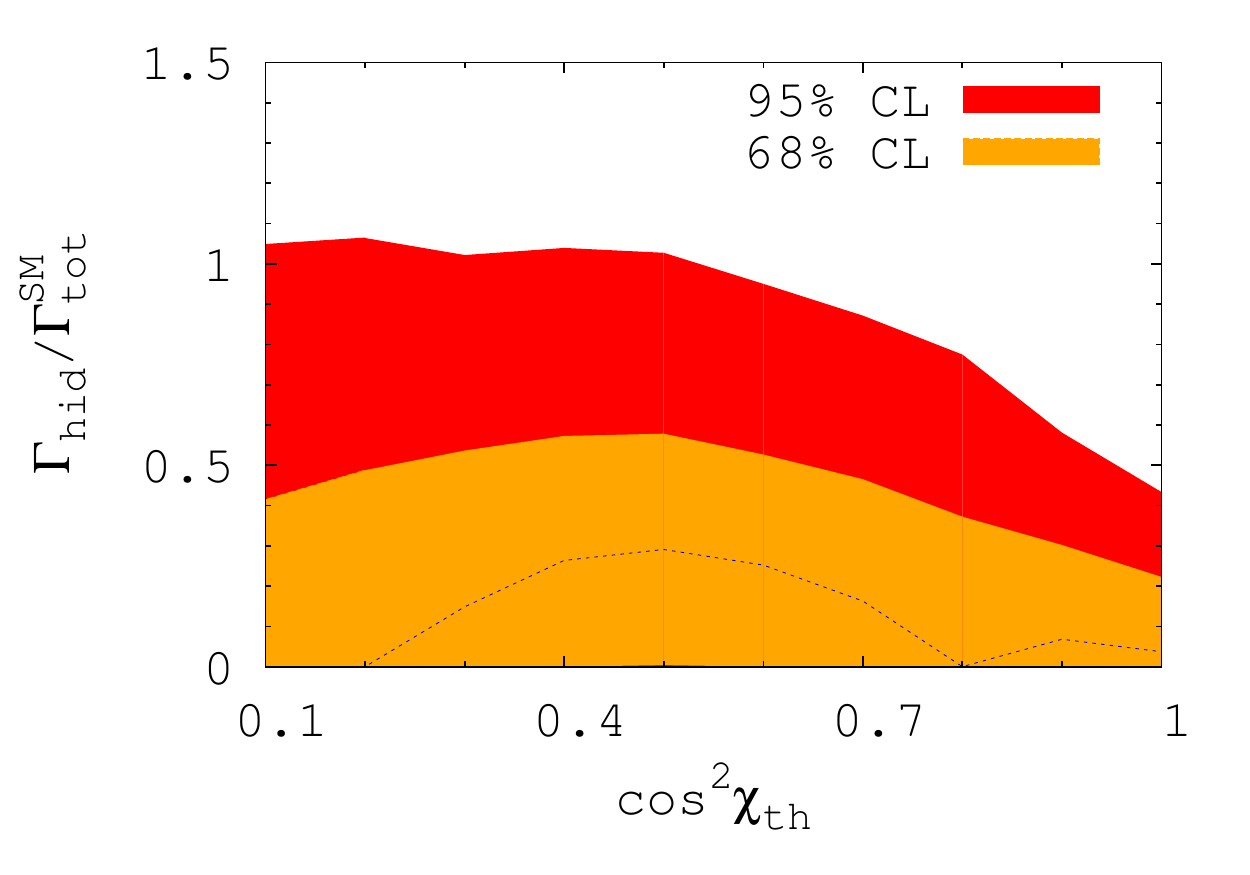}
\includegraphics[width=0.36\textwidth]{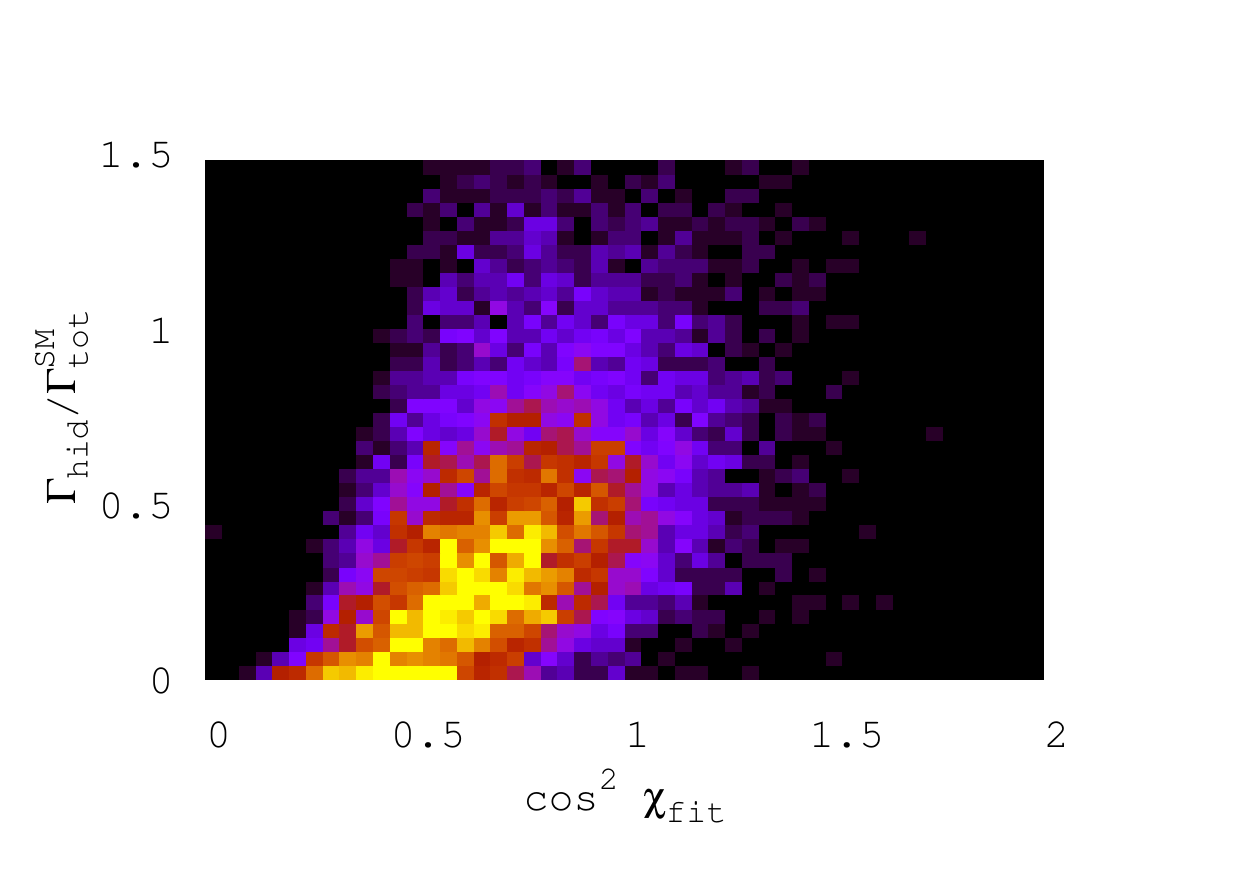} 
\end{center}
\vspace*{-8mm}
\caption{LHC sensitivity to modified Higgs couplings and invisible
  decays, based on 30~fb$^{-1}$ of data. Upper row: $\Gamma_\text{hid} 
  = 0$; lower row $\Gamma_\text{hid} = \sin^2\chi \, 
  \Gamma^\text{SM}_{\text{tot}}$ for invisible decays. 
  The Higgs mass is fixed to 120~GeV.
  Left column: extracted $\cos^2 \chi_\text{fit}$ values as a function 
  of $\cos^2 \chi_\text{th}$; Center column: extracted bounds and
  measurements of   $\Gamma_\text{hid}/\Gamma^\text{SM}_\text{tot}$ 
  as a function of $\cos^2 \chi_\text{th}$; Right column: illustration 
  of the correlation between mixing and invisible partial width using
  $\cos^2 \chi_\text{th} = 1.0$ [upper row] and $0.6$ [lower row]. 
}
\label{fig:portal_inv}
\end{figure}

Before moving away from Standard Model decays only, 
we show the results of a two-parameter fit
in the upper panels of Figure~\ref{fig:portal_inv}, while 
keeping $\Gamma_{hid} = 0$. Compared to the
one-dimensional parameter extraction shown in
Figure~\ref{fig:portal_noinv} the error on the extracted value of
$\cos^2 \chi$ is now increased. 
This arises because we now have to perform a
two-dimensional parameter extraction and project the correlated
uncertainty onto the two model parameters $\cos \chi$ and
$\Gamma_\text{hid}$. Comparing the upper panels of
Figure~\ref{fig:portal_inv} to Figure~\ref{fig:portal_noinv} shows this
effect on the extracted parameters, just fitting two parameters
without even introducing an invisible Higgs decay. The error on
$\cos^2 \chi$ becomes asymmetric due to the positivity constraint on
the Higgs width, fixing the upper error band of $\cos^2 \chi$. On the
other hand, fake invisible-decay events from background fluctuations
lead to a wide lower error bound.
Moreover, we observe a clear bias towards
too large event numbers towards small values of $\cos^2 \chi$. This is
in part due to the asymmetric Poisson distribution and in part due to
the fact that measurement channels where the number of background
events from the control region exceeds the number of events in the
signal region, are explicitly excluded from the fit. A more detailed
discussion of this effect is presented in
Appendix~\ref{sec:appobsbias}. Finally, the error band for the extracted
$\Gamma_\text{hid}$ ranges around 20\% of the Standard Model width.

The impact of actual invisible decays is estimated by choosing
$\Gamma_{\text{hid}} = \sin^2\chi \, \Gamma^{\text{SM}}_{\text{tot}}$
as illustrative example for the parametrization of the non-zero
invisible partial width. This parametrization accounts naturally for
the suppression of $\Gamma_\text{hid}$ by mixing, while a coefficient
of size $\Gamma^{\text{SM}}_{\text{tot}}$ would be expected for
structures in the hidden sector roughly parallel to the standard
sector~\cite{higgs_portal}. Comparing the upper and lower left panels of
Figure~\ref{fig:portal_inv}, it is evident that the effect of an actual
invisible Higgs decay on the extraction of $\cos^2 \chi$ is small.  A
measurement of the invisible Higgs width, beyond setting upper bounds,
seems challenging with only 30~fb$^{-1}$ of integrated luminosity. The
best discrimination power we obtain for medium-sized values of $\cos^2
\chi$, where we have a significant invisible branching ratio and the
production side is not strongly suppressed.

Finally, in the right panels we see a clear correlation between the
two extracted parameters, owed to the form of the observables shown in
Eq.(\ref{eq:obs}) and Figure~\ref{fig:corr}.

\section{Strongly Interacting Higgs Boson}
\label{sec:strong}

Deviations of the Higgs couplings similar to the $\cos \chi$ factor in Eq.(\ref{eq:paras}) are 
expected when a light
Higgs boson is generated as a pseudo-Goldstone boson by global symmetry
breaking in a new strong interaction sector. Depending on the details
of the model, the Higgs couplings are modified either
individually for different particle species, or universally for all
species~\cite{silh,silh_pheno,MMMM}. In contrast to the hidden Higgs model, the
light Higgs boson $H_1$ does not decay into channels
not present in the Standard Model.

\begin{figure}[t]
\begin{center}
\includegraphics[width=0.32\textwidth]{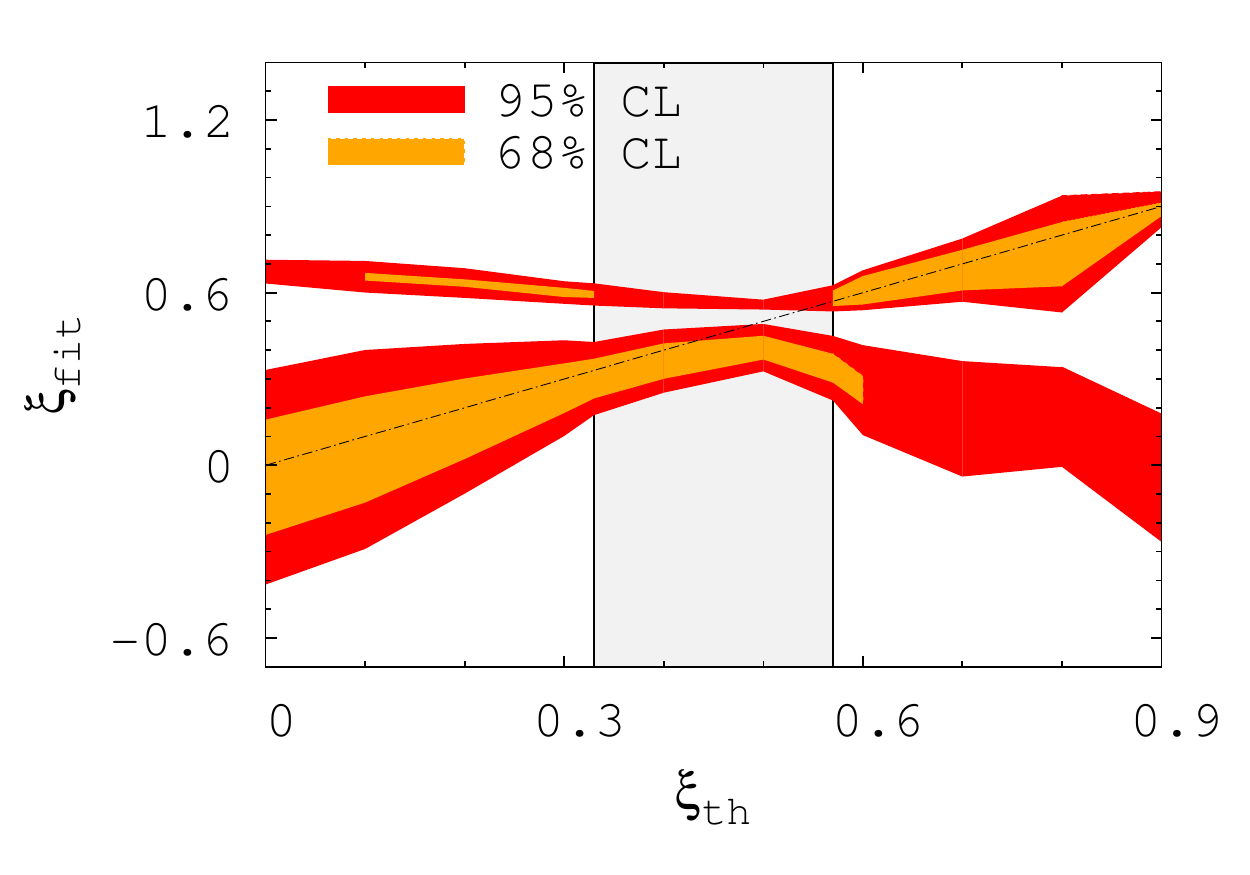}
\includegraphics[width=0.32\textwidth]{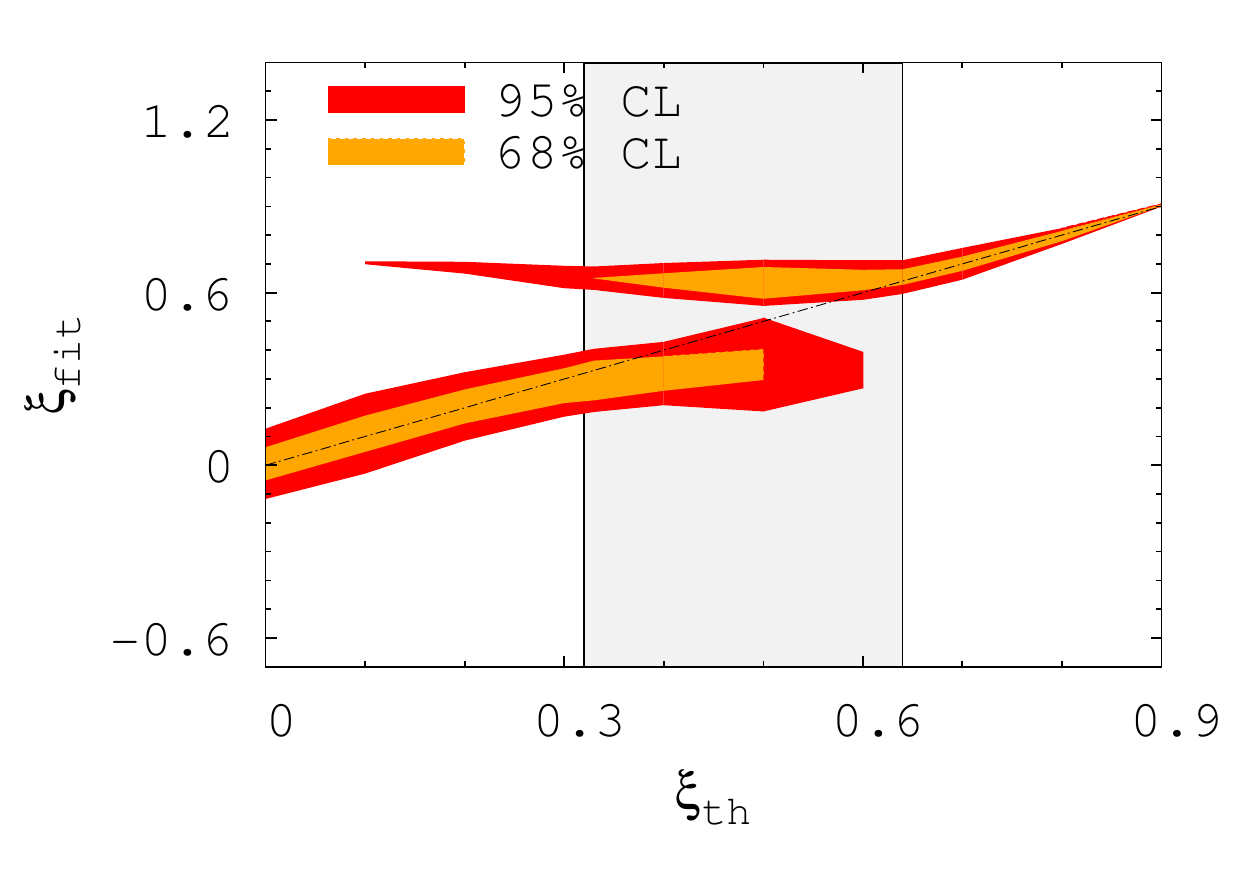}
\includegraphics[width=0.32\textwidth]{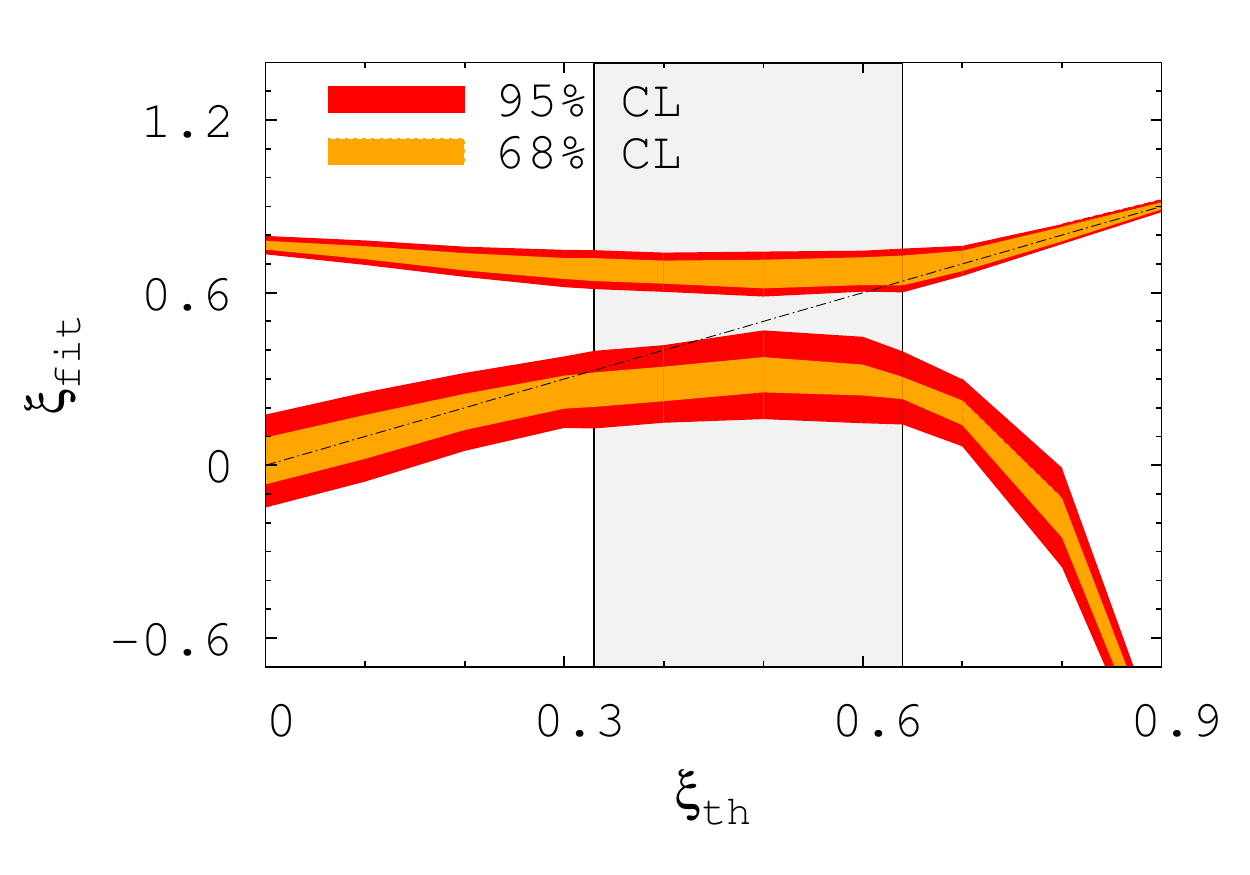} \\[-2mm]
\includegraphics[width=0.32\textwidth]{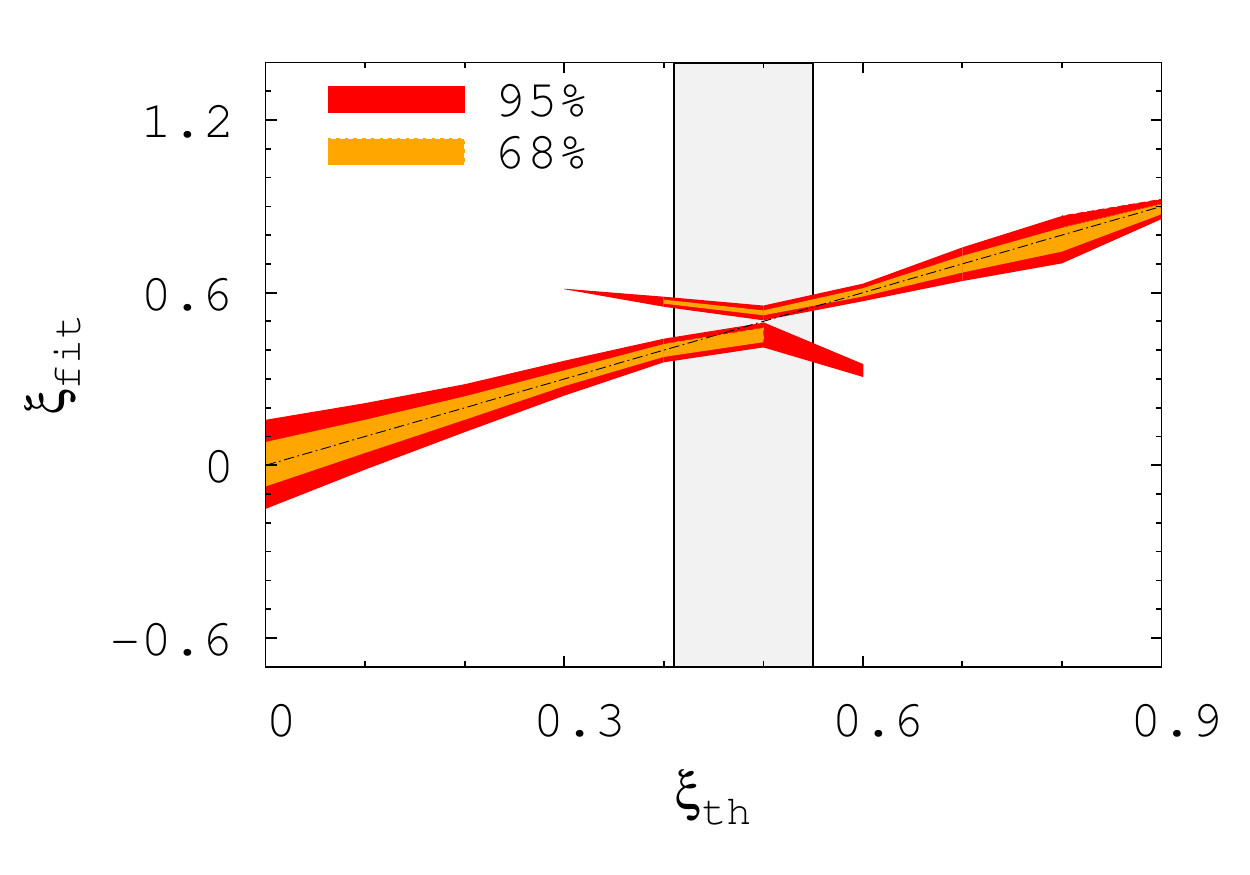}
\includegraphics[width=0.32\textwidth]{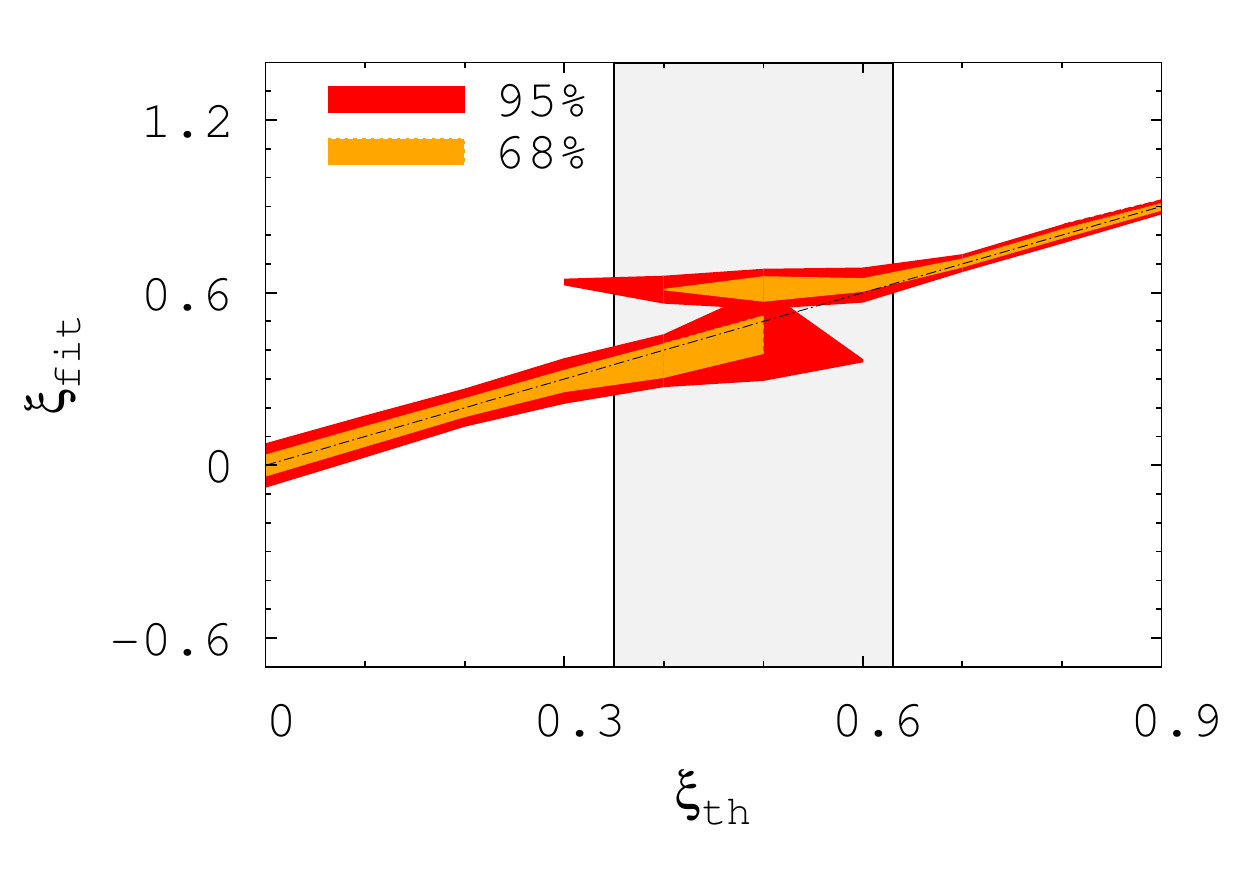}
\includegraphics[width=0.32\textwidth]{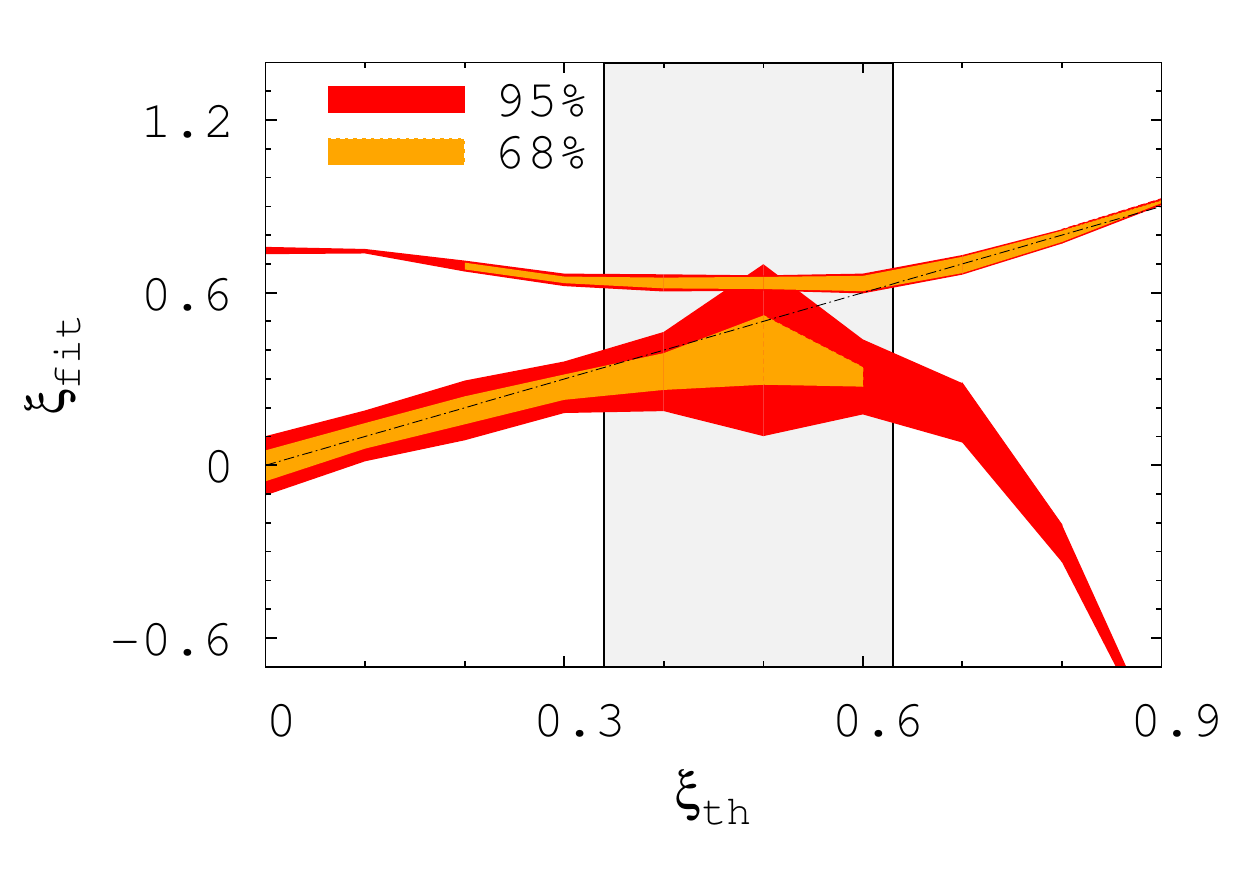}
\end{center}
\vspace*{-8mm}
\caption{LHC sensitivity to modified Higgs couplings 
  based on $30$ [upper] and 300~fb$^{-1}$ [lower row] for un-aligned boson
  and fermion couplings
  as a function of the assumed $\xi_\text{th}$ for $m_H=120$~GeV 
  (left), 160~GeV (center) and 200~GeV (right). $\xi$ values 
  close to 1/2, for which the rates are strongly suppressed, are
  blinded by the gray bars.}
\label{fig:maggi5}
\end{figure}

Such a picture can be developed using Holographic
Higgs Models, based on the AdS/CFT correspondence, in which strongly
coupled theories in four dimensions are identified with weakly coupled
theories in five dimensions~\cite{holographic}. For example, the spontaneous breaking of a global symmetry  
$SO(5) \to SO(4)$ generates the adequate iso-doublet of Goldstone bosons. Assigning the Standard Model fermions
either to spinorial or fundamental $SO(5)$ representations changes
the Higgs couplings 
either universally or separately for Standard Model vectors and
fermions{\footnote{With $f$ larger than $v$, the mass scale of the
new fields in such scenarios will be about 1~TeV and above. Their
effect on Higgs couplings to SM fields through loops will therefore,
quantum mechanically, be strongly suppressed. Though the normal SM
modes are largely decoupled near $\xi = \frac12$, {\it cf.} Appendix B,
the overall coupling mediated by the new fields will remain small
and the production of the Higgs particles in the parameter range
will correspondingly be suppressed. In fact, this parameter range must
be blinded in our analysis as a result, {\it cf.}
Figure~\ref{fig:maggi5}}}. The modifications are determined by the
parameter 
\begin{equation}
\xi = \left( \frac{v}{f} \right)^2
\end{equation}
which measures the magnitude of the Goldstone scale $f$ in relation to the
standard Higgs vacuum expectation value $v$.  
The case where all Higgs
couplings are suppressed universally by a factor $(1-\xi)^{1/2}$, 
is covered by the analysis in the preceding section
identifying
\begin{equation}
\kappa \equiv 1-\xi   \,.
\end{equation}
All results from the Higgs portal can be transferred
one-by-one to this universal strong interaction model, identifying
$\cos^2 \chi \to 1 - \xi$ and setting $\bratio = 0$. Hence, 
it can be concluded from Figure~\ref{fig:portal_noinv} that $\xi$ can be measured 
with an uncertainty of $10\% - 20\%$, depending on the Higgs mass.
\medskip

%
In a closely related scenario~\cite{MMMM}
universality is broken to the extent that the Higgs coupling
of vector particles is reduced, still, by $(1-\xi)^{1/2} \approx 1 -
\xi/2$ but the coupling of fermions by a different coefficient,
$(1-2\xi)/(1-\xi)^{1/2} \approx 1-3\xi/2$.  Now, the two $\kappa$ parameters of
the twin width-ratios, at small $\xi$, read
\begin{alignat}{5}
\kappa_V &= 1- \left[ 1 (3) -2 \, \brsm_{\text{f}} \right] \, \xi     \notag \\
\kappa_f &= 1- \left[ 3 (5) -2 \, \brsm_{\text{f}} \right] \, \xi     \,,
\end{alignat}
The indices $V,f$ distinguish vector and fermion Higgs decays, and the expression in brackets corresponds to production either in  
Higgs-strahlung/electroweak boson fusion or --- altered to () ---
in gluon fusion; $\brsm_{\text{f}}$ denotes the inclusive Higgs branching ratio 
to fermions in the Standard Model which 
is close to one for light Higgs bosons. 

The two $\kappa$ parameters of this non-universal strong interaction model
are characteristically different from both the universal strong 
interaction model as well as the Higgs portal. First, they are 
different for vector and fermion decays of the Higgs particle; 
second, in parameter regions in which more than half of the Higgs
decays are fermionic, $\kappa_V$ is larger than unity for 
Higgs-strahlung/electroweak fusion.
\medskip 

%
\begin{table}[t]
\begin{tabular}{|c|c||c|c|c|c|c|c||c|c|c|c|c|c|}
\hline
\multicolumn{2}{|c||}{}
              & \multicolumn{6}{c||}{$30\text{ fb}^{-1}$}
              & \multicolumn{6}{c|}{$300\text{ fb}^{-1}$} \\
\hline
$M_H$ [GeV] & $\;\;\xi_{\text{th}}\;\;$ 
            & $\;\;\xi_1\;\;$ & $\;\;\Delta\xi_1\;\;$ & $\;\;w_{\xi_1}\;\;$
            & $\;\;\xi_2\;\;$ & $\;\;\Delta\xi_2\;\;$ & $\;\;w_{\xi_2}\;\;$
            & $\;\;\xi_1\;\;$ & $\;\;\Delta\xi_1\;\;$ & $\;\;w_{\xi_1}\;\;$
            & $\;\;\xi_2\;\;$ & $\;\;\Delta\xi_2\;\;$ & $\;\;w_{\xi_2}\;\;$ \\
\hline\hline
120& 0.0 &-0.04 & 0.36 & 0.94 & 0.67 & 0.06 & 0.06 & 0.0  & 0.14 & 0.99 & 0.67 & 0.08 & 0.01 \\
   & 0.2 & 0.16 & 0.26 & 0.88 & 0.63 & 0.06 & 0.12 & 0.20 & 0.08 & 0.99 & 0.63 & 0.02 & 0.01 \\
   & 0.6 & 0.61 & 0.06 & 0.54 & 0.26 & 0.18 & 0.46 & 0.60 & 0.02 & 0.96 & 0.31 &  0.04 & 0.04 \\
\hline
160& 0.0 & 0.0  & 0.12 & 0.98 & 0.73 & 0.03 & 0.02 & 0.0  & 0.08 & 1.0  & ---  & ---  & 0    \\
   & 0.2 & 0.20 & 0.10 & 0.95 & 0.69 & 0.04 & 0.05 & 0.20 & 0.08 & 0.99 & 0.67 & 0.04 & 0.01 \\
   & 0.6 & 0.64 & 0.06 & 0.85 & 0.33 & 0.10 & 0.15 & 0.63 & 0.04 & 0.96 & 0.36 & 0.08 & 0.04 \\
\hline
200& 0.0 & 0.01 & 0.16 & 0.84 & 0.77 & 0.03 & 0.16 & 0.0  & 0.10 & 0.92 & 0.75 & 0.014& 0.08 \\
   & 0.2 & 0.19 & 0.14 & 0.67 & 0.71 & 0.04 & 0.33 & 0.19 & 0.10 & 0.79 & 0.69 & 0.018& 0.21 \\
   & 0.6 & 0.67 & 0.06 & 0.50 & 0.30 & 0.16 & 0.50 & 0.63 & 0.03 & 0.55 & 0.31 & 0.16 & 0.45 \\
\hline
\end{tabular}
\caption{Errors $\Delta\xi$ [95\% CL] on the pseudo-Goldstone parameter $\xi$
         for integrated luminosities of 30 [left] and 
         300 fb$^{-1}$ [right]. Shown are the two solutions together with their
         corresponding probability $w_{\xi}$ of the best-fit, 
	 which are the relative numbers of toy experiments ending up in
	 the vicinity [as obtained by a fit of two Gauss peaks] of this
         solution. }
\label{tab:xsi}
\end{table}

Results for this model are presented in Figure~\ref{fig:maggi5} and
in Table~{\ref{tab:xsi}} for 30 and 300~fb$^{-1}$.  At low luminosity two
solutions emerge, while an increased luminosity eliminates the fake solution 
in major parts of the parameter space, as long as the Higgs mass is small.
The 
mathematical evolution of the two solutions and their analytical form is discussed
in Appendix~\ref{sec:appaltsol}.
Around $\xi \sim 0.5$ the observable rates at the LHC drop sharply,
not allowing for a reliable extraction of $\xi$ simply based on 
too little statistics for the fit.
Therefore, the gray bands blind
ranges of parameter space in which the modified theory leads to 
too large a suppression. Only when our fit finds $\chi^2/\text{d.o.f}
\gtrsim 1$ the result becomes statistically
trustworthy again. 

In contrast to 120~GeV, we observe that the situation hardly
improves for a 200~GeV Higgs boson once we go to higher luminosity. 
For this mass only four
Higgs channels are left. All include a decay into either $W$ or $Z$
bosons, and on the production side three of them are proportional to
$g_{ttH}$, either via gluon-fusion or top-quark associated
production. These three measurements are equivalent and exhibit an
ambiguity because fermion-Higgs couplings cannot distinguish between
small and large values of $\xi$, see Appendix~\ref{sec:appaltsol}.  
Only the fourth measurement based on
weak-boson fusion production can resolve the ambiguity.  Even though
this channel is comparably clean, it is systematics limited, \ie it
hardly improves with higher luminosities.

\begin{figure}[t]
\begin{center}
\hspace*{-3ex}
\includegraphics[width=0.36\textwidth]{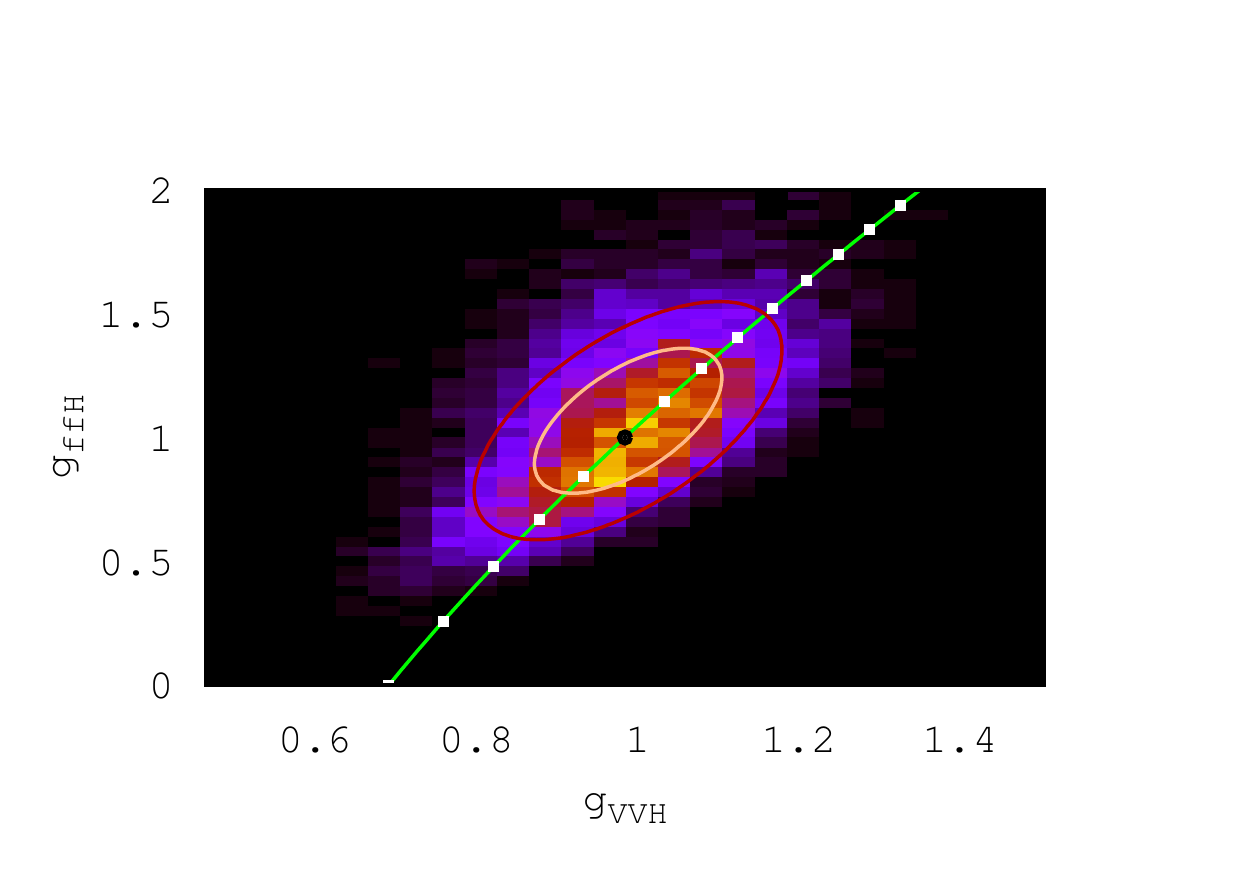}
\hspace*{-6ex}
\includegraphics[width=0.36\textwidth]{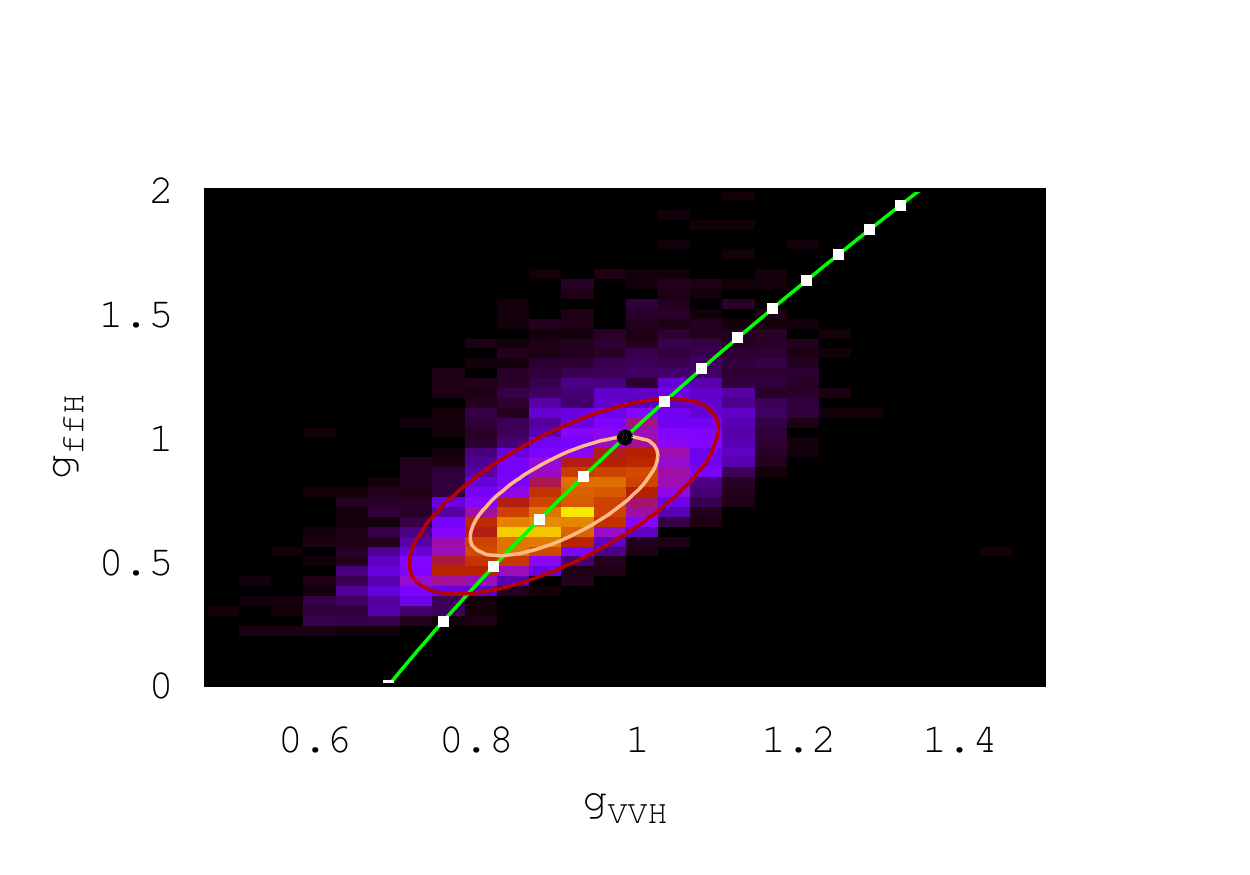}
\hspace*{-6ex}
\includegraphics[width=0.36\textwidth]{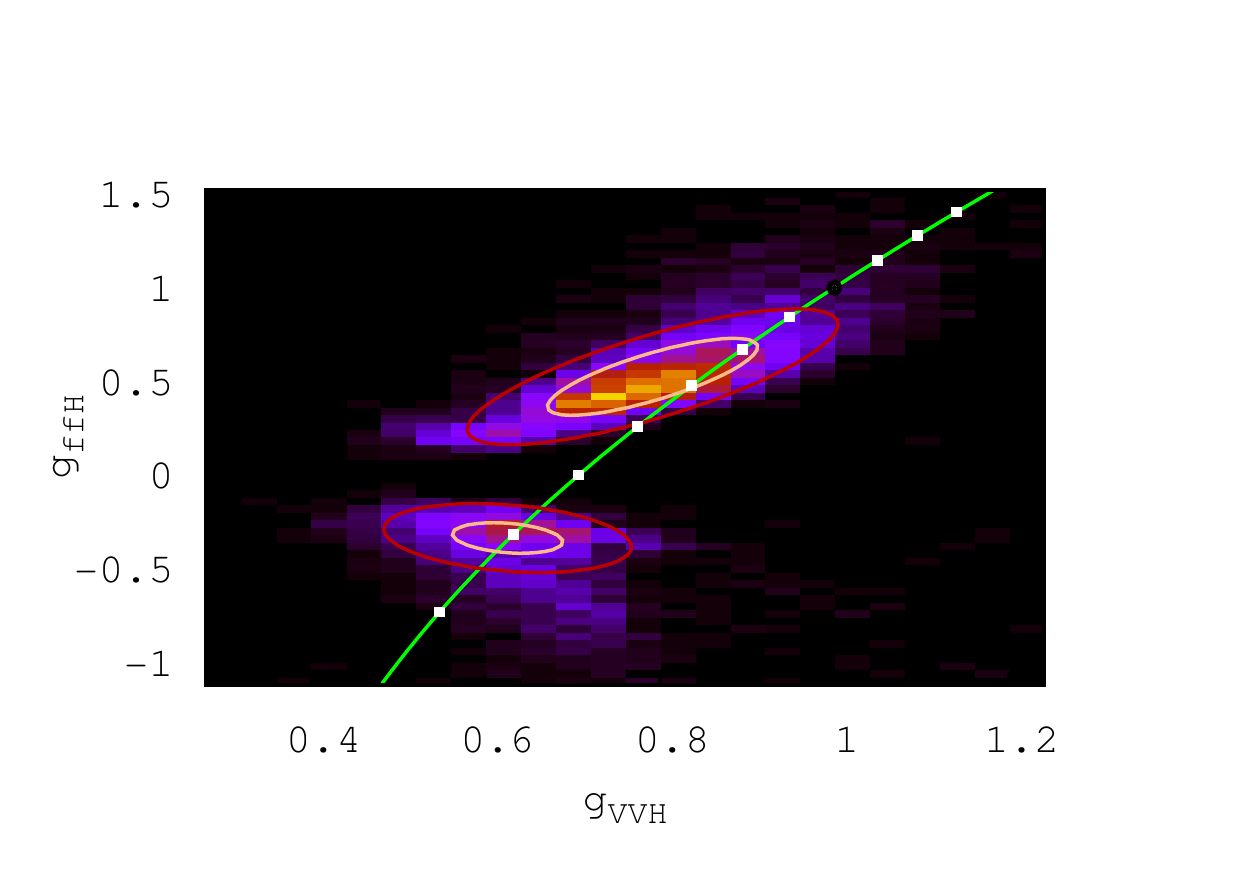}
\hspace*{-6ex}
\end{center}
\vspace*{-8mm}
\caption{Independent fit of common vector [$g_{VVH}$] and fermion
[$g_{ffH}$] Higgs couplings for $\xi_\text{th}=0$, $0.2$ and $0.6$ using
30~fb$^{-1}$. The $\xi$ values correspond to $g_{VVH} = 1$, $0.89$,
$0.63$ and $g_{ffH} = 1$, $0.67$, $-0.32$, respectively.
The red and orange ellipses show the 68\% and 95\% CL regions.
The green line marks the model prediction, with the black dot indicating
$\xi=0$ and the white dots in 0.1 distance, moving to positive values
towards the lower left.
}
\label{fig:gv_gf}
\end{figure}
Also we need to establish that the couplings indeed follow this pattern.
To quantify this we perform a fit of vector and fermion couplings
separately, illustrated in Fig.~\ref{fig:gv_gf} using theory 
scenarios for $\xi_{\text{th}}=0$ (SM), $0.2$ and $0.6$ with a luminosity of
30~fb$^{-1}$. We find for the SM case that we can determine the vector
couplings with a $2\sigma$ error of 0.22 and the fermion couplings with
an accuracy of 0.55. A positive correlation between the two parameters
of 0.59, which is approximately aligned with the model prediction as
function of $\xi$, further improves our possibility to exclude
deviations. The 95\% CL region intersects the theory line at $\xi =
-0.38$ and $0.24$.
For $\xi_{\text{th}}=0.2$ the changes are small, with a slightly
improved precision on the fermionic couplings and a larger
correlation. In the $\xi_{\text{th}}=0.6$ case the fake solution around
smaller $\xi$ values is again similar, while the correct one yields a
different behavior. 
The correlation has changed sign, making it more difficult to
discern it from other models which do not follow the distinct pattern.
As two standard deviation errors we obtain smaller values for this
solution on the other hand. They are  $0.16$ and $0.20$ for fermion and
vector couplings, respectively.
\medskip

The two solutions shown in Figure~\ref{fig:maggi5} also represent a
technical challenge. We need to know not only the error $\Delta\xi$ on
the individual solution, but also the probability $w_{\xi}$ with which
this solution appears. A Minuit fit per toy-experiment alone cannot
achieve this, because there is no guarantee that it will find the
global minimum of $\chi^2$. Indeed, we observe a strong dependence on
the starting point in Minuit. Therefore, we need to add a coarse grid
scan of the one-dimensional parameter space before the minimization.
To obtain the $68\%$ and $95\%$ confidence levels in
Figure~\ref{fig:maggi5} we first fit a sum of two Gaussians to the
parameter distribution. To extract a confidence level interval around
each of the two peaks we define a height which crosses both Gaussians
such that the sum of the two central areas corresponds to $68\%$ or
$95\%$ of the entire integral under both Gaussians. The two
intersections of this horizontal line with each Gaussian give us the
confidence regions quoted for example in Figure~\ref{fig:maggi5}.

First of all, in Figure~\ref{fig:maggi5} we again see that the typical
error bars shrink when we increase the Higgs mass and with it the
number of events from 120~GeV to 160~GeV. For even larger Higgs masses
some of the relevant channels rapidly vanish, so the error bars
increase again. The actual error bars for the different scenarios are
listed in Table~\ref{tab:xsi}. In particular for moderate luminosity
the absolute error on $\xi_1$, for assumed values of $\xi_\text{th} <
0.2$ range around 20\% for a light Higgs boson and well below 10\%
towards larger mass. These numbers again roughly compare to the
typical values for $\Delta \xi$ in the universal strongly interacting
model or $\sqrt{\Delta \kappa}$ for the Higgs portal. For larger
values of $\xi \geq 0.6$ the two strongly interacting models start
deviating significantly. In particular, the relative error $\Delta
\xi_1/\xi_1$ decreases dramatically. The reason will be discussed in
detail in Appendix~\ref{sec:appaltsol}: due to the non-universal
structure of the fermion and vector couplings in this range the Higgs
event rates vary much more strongly than elsewhere. Therefore, the LHC will
probe this region with high precision, with the slight caveat that
towards slightly smaller $\xi$ the Higgs production and decay rates
vanish rapidly. These are precisely the dangerous gray regions in
Figure~\ref{fig:maggi5} --- as long as we observe Higgs events at the
LHC the sharp drop in rates towards $\xi=0.5$ is a welcome feature to
analyze the model at colliders.
\medskip

The improvement of the errors with increased statistics is also shown in
Table~{\ref{tab:xsi}}.  If the measurement were statistics dominated, the error
would be reduced by a factor $\sqrt{10}\simeq3$ when going from 30 to
300~fb$^{-1}$. From the results we find that the improvement is
smaller. The reason is that all measurements determine only
a single parameter, so unlike in the unconstrained Higgs sector
analysis~\cite{sfitter_higgs} in the current setup even for 30~fb$^{-1}$
the results are not entirely statistics dominated.

\section{Summary}
\label{sec:outlook}

In this letter we have analyzed the LHC reach for deviations of Higgs
couplings from their Standard Model values. Two models can be linked
to such signatures: First, we couple the Standard Model Higgs field to
a hidden Higgs sector, opening a renormalizable Higgs portal between
the two sectors.  From the event rates for visible Higgs production
and decay channels we could derive upper bounds on non-SM
admixtures in the wave-function of the Higgs boson and on novel
invisible decay channels. Since it is unclear if at the LHC we will be
able to quantitatively analyze invisible Higgs decays, given the
experimental errors, the question arises if we include the invisible
Higgs decay rate in a two-dimensional fit or only include modified
visible couplings. 

From the two-dimensional fit we find 95\% CL limits on the two model
parameters $\sin^2\chi < 0.50$ and $\Gamma_\text{hid}/\gtot^\text{SM}
< 1.0$ ($\mathcal{L} = 30~\ifb$ and $m_H = 120$~GeV).  Towards Higgs
masses around 160~GeV these limits slightly improve, while for even
heavier masses the decreasing production rates at the LHC take their
toll. These numbers correspond to a $10\%$ to $20\%$ measurement of
the modified Higgs couplings $(\Delta g)/g$.  Such bounds also quantify to
what accuracy a Higgs boson discovered and studied at the LHC could be
identified as the Standard Model realization. They should be compared
to limits on the individual Higgs couplings from general Higgs sector
analyses~\cite{duehrssen,sfitter_higgs} and they show a significant
improvement driven by the higher level of specification in the
hypothesis tested.

Neglecting the invisible Higgs decay and hence trading the minimal
amount of extra information for the advantage of a one-dimensional fit
without additional noise in a small region of parameter space ($m_H
\sim 160$~GeV and $\cos^2 \chi \sim 0.6$) we might even be able to
establish a Higgs portal at the $5\sigma$ level based on an integrated
luminosity of 30~fb$^{-1}$.

If invisible decay channels of the Higgs boson are indeed measured,
the mixing parameter and the partial width for Higgs decays to the
hidden sector can of course be determined individually. However, such
a measurement will likely require higher luminosities.
\medskip

A related problem arises when a light Higgs boson is identified with a
pseudo-Goldstone boson associated with the spontaneous breaking of
global symmetries in new strong interactions. If fermion and vector
couplings are scaled equally, the scaling analysis of the hidden
sector can be transcribed without change, assuming there are no Higgs
decays into non-SM channels.  The bound on $\sin^2\chi$ can simply be
read as a bound on $\xi$.

More interestingly, if the Higgs couplings are modified separately for
vectors and fermions, various production/decay channels may be
analyzed individually so that a unique picture emerges. For small
values of $\xi$ the two kinds of couplings at least qualitatively
scale similarly, so our results for $(\Delta \xi)/\xi$ follow the
usual patterns. For larger values around $\xi \sim 0.5$ the
Higgs-fermion couplings now vanish, leading to wells in the Higgs
event rates. In those regions the relative error on the determination
of $\xi$ can shrink to 5\%, though being dangerously close to
parameter regions where the Higgs discovery would require larger LHC
luminosities.\medskip

The LHC will be able to probe scenarios with
modified Higgs couplings as generally analyzed in 
Refs.~\cite{duehrssen,sfitter_higgs}. However, testing a specific
one- or two-parameter model appears to be a promising strategy to gain
insight into the Higgs sector already based on an integrated
luminosity of 30~fb$^{-1}$ at 14~TeV. Compared to the general analysis
the typical error bars on Higgs couplings are reduced by at least a
factor $1/2$, now ranging around $10\%$ to $20\%$ for Higgs masses
between 120~GeV and 200~GeV.


\acknowledgments{}

PMZ is grateful to the Institut f\"ur Theoretische Teilchenphysik und
Kosmologie for the warm hospitality extended to him at RWTH Aachen
University.  MR acknowledges support by the Deutsche
Forschungsgemeinschaft via the Sonderforschungsbereich/Transregio
SFB/TR-9 `Computational Particle Physics' and the Initiative and
Networking Fund of the Helmholtz Association, contract HA-101
(Physics at the Terascale). MR, DZ and PMZ are grateful to the Institut
f\"ur Theoretische Physik of Heidelberg University for the
hospitality. Part of the work by DZ is supported by the GDR Terascale of the CNRS.

\appendix

\section{Observational bias}
\label{sec:appobsbias}

Statistical fluctuations in the measurements can
lead to situations which cannot be interpreted in terms of physically
meaningful parameters. One such example is a negative number of signal
events. At the LHC we first measure the sum of signal and background
events in the signal region. The number of background events we then
determine either from extrapolation from a signal-free control region,
like sidebands, or from Monte Carlo simulations. The difference gives
us the number of signal events. At arbitrarily large statistical
significances the difference between the two measurements, \ie the
number of signal events, is by definition much larger than the
statistical error on each of them individually. However, for
significances between one and three standard deviations as we expect
them for the Higgs sector at the LHC a downward fluctuation in
signal-plus-background and an upward fluctuation in background-only
can lead to a negative difference. This effect has to be dealt with
when we compose the sample of measurements which for example enter the
Higgs sector analysis presented in this paper. The question is, if
solutions to this problem will affect for example the central values
and errors quoted for the Higgs couplings.\medskip

\begin{figure}[b]
\begin{center}
\includegraphics[width=0.3\textwidth]{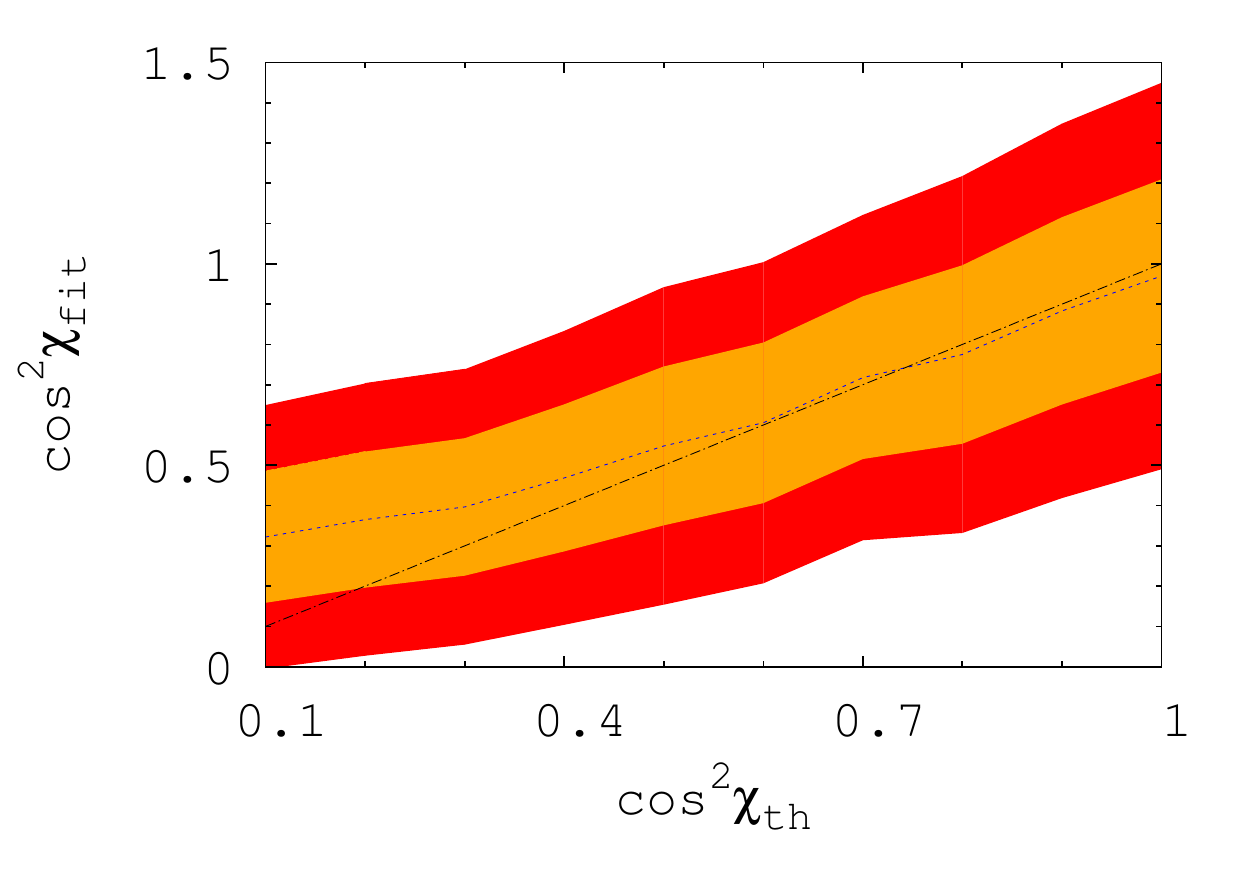}
\includegraphics[width=0.3\textwidth]{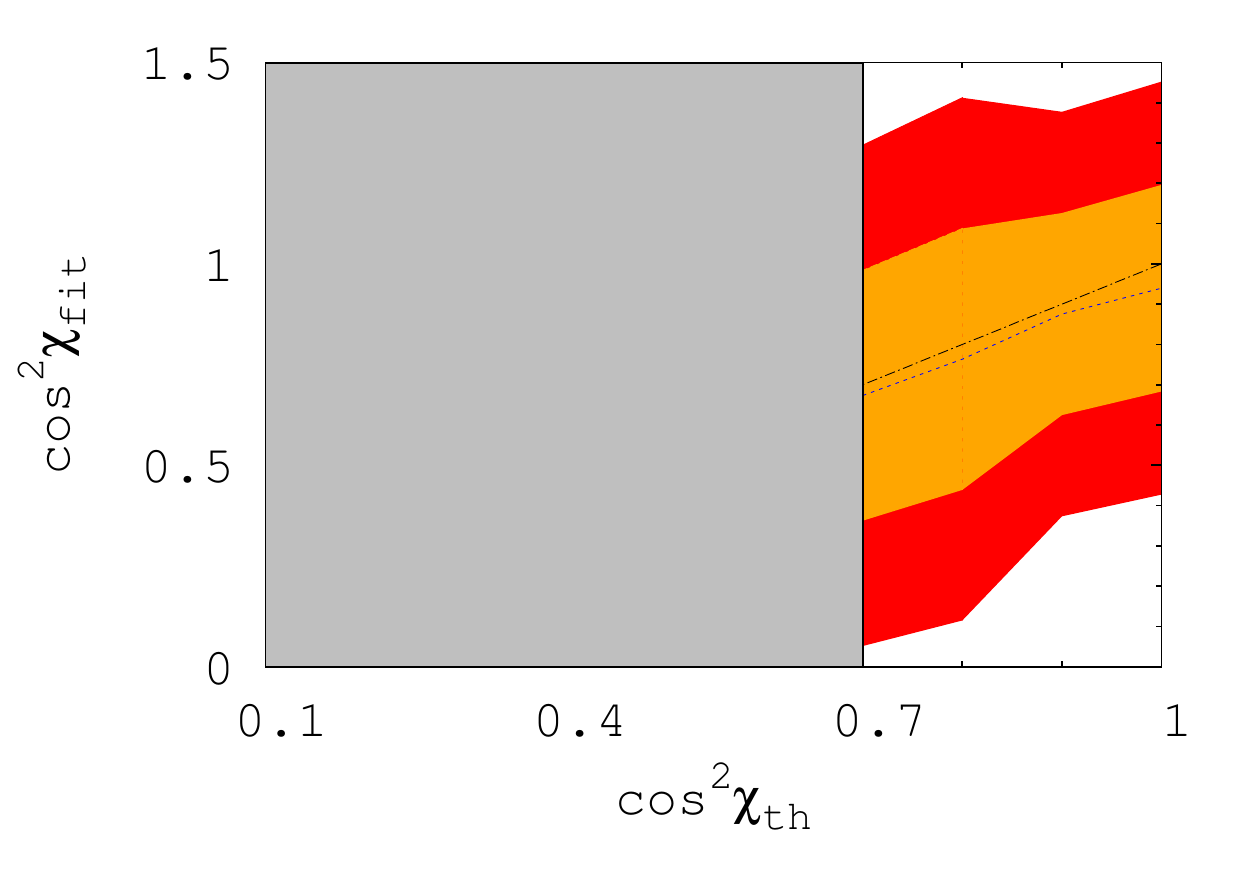}
\includegraphics[width=0.3\textwidth]{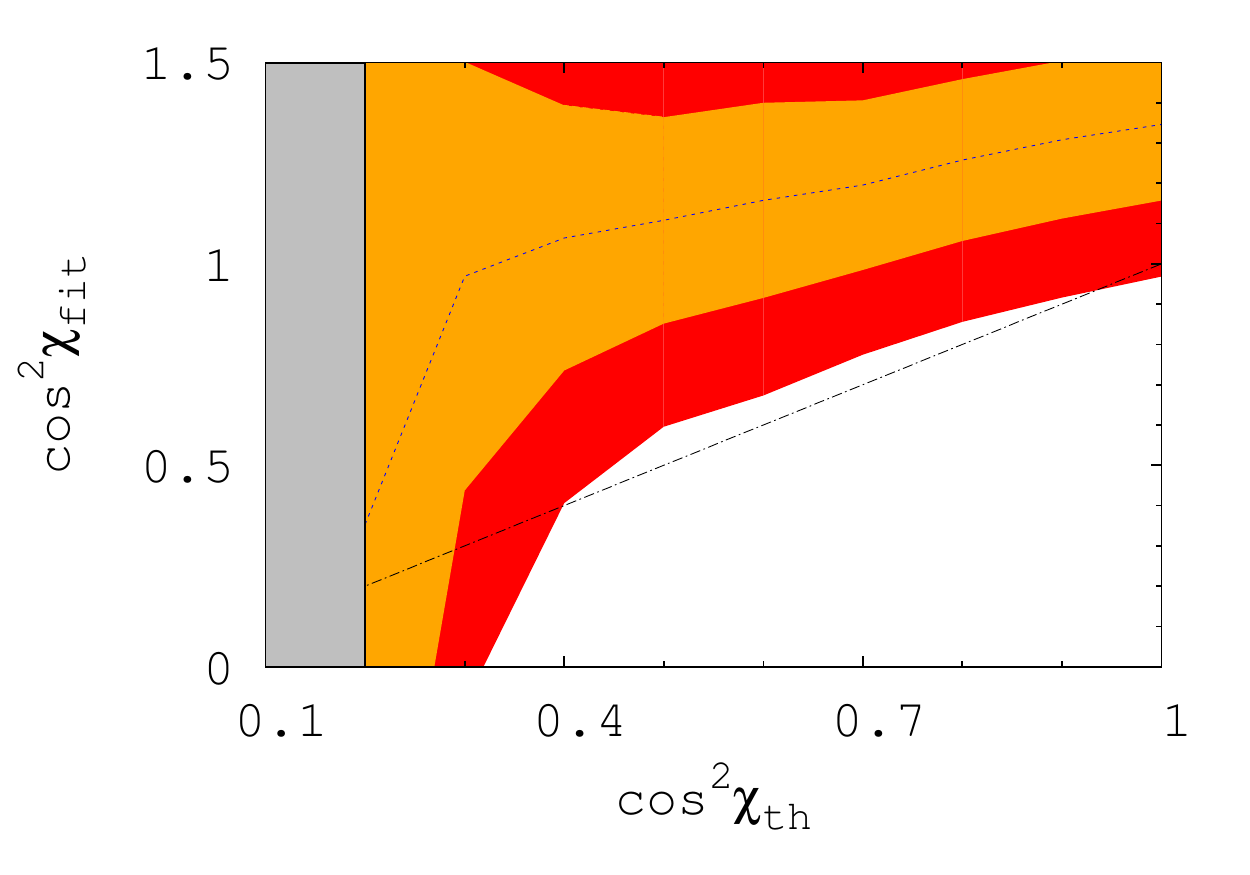}
\end{center}
\vspace*{-8mm}
\caption{Different treatment of channels, depending on the number of
  events in the signal region $S+B$, compared to background events $B$
  from a signal-free control region.  Left: a measurement is included
  when $S+B > B$.  Center: only measurements corresponding to at least
  a $2\sigma$ excess for nominal signal and background rates are
  included.  Right: only measurements for which the measured values of
  $S+B$ and $B$ give at least two standard deviations are included.  }
\label{fig:alttreat}
\end{figure}

We present three alternative treatments in Figure~\ref{fig:alttreat}. On
the left-hand side we only take into account channels where we measure
a positive number of signal events. This prescription we use in our
analysis. For couplings fairly close to their Standard Model values we
obtain the correct central value. For small couplings we observe a
significant shift to larger values.  The reason is that if a
measurement shows an upward signal fluctuation we include it, while a
downward fluctuation quickly reaches the negative-$S$ threshold and
gets excluded.

In the central panel we only include measurements where the nominal
number of signal and background events yields a $2 \sigma$ excess. We
find good agreement between fitted and truth values of $\chi$, but for
$\cos^2\chi < 0.7$ there are no measurements left which fulfill this
condition.

For the right panel we only include channels where the measured
numbers yield a $2 \sigma$ excess. Here, we observe a significant
upward shift over the whole parameter range. Even for Standard Model
couplings the actual Standard Model is almost excluded at the 95\% CL.
Therefore, it is important to take into account measurements which
have a low observed significance on their own. They can provide upper
bounds on Higgs couplings and avoid a bias generated by upward
fluctuations of signal events in some of the channels included.

\begin{figure}[t]
\begin{center}
\includegraphics[width=0.40\textwidth]{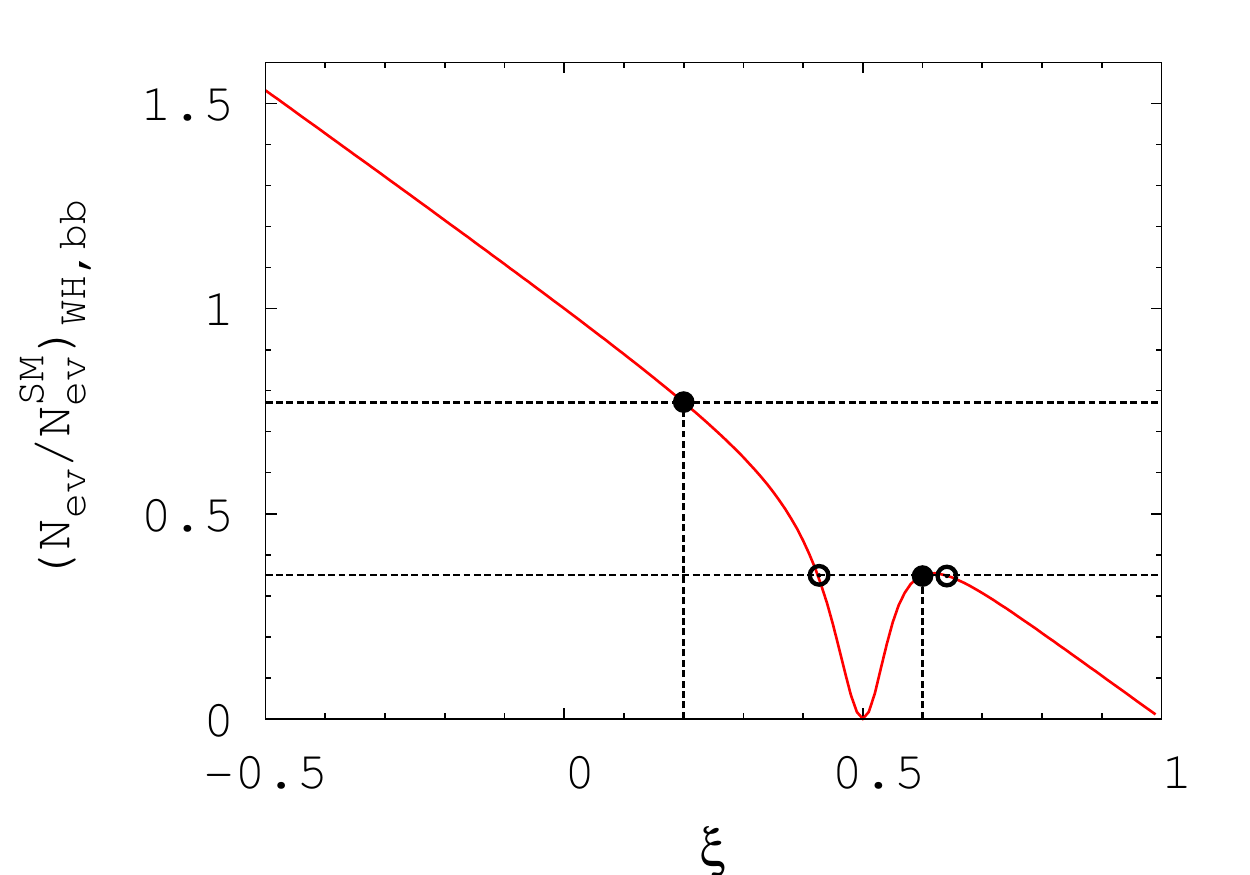}
\includegraphics[width=0.40\textwidth]{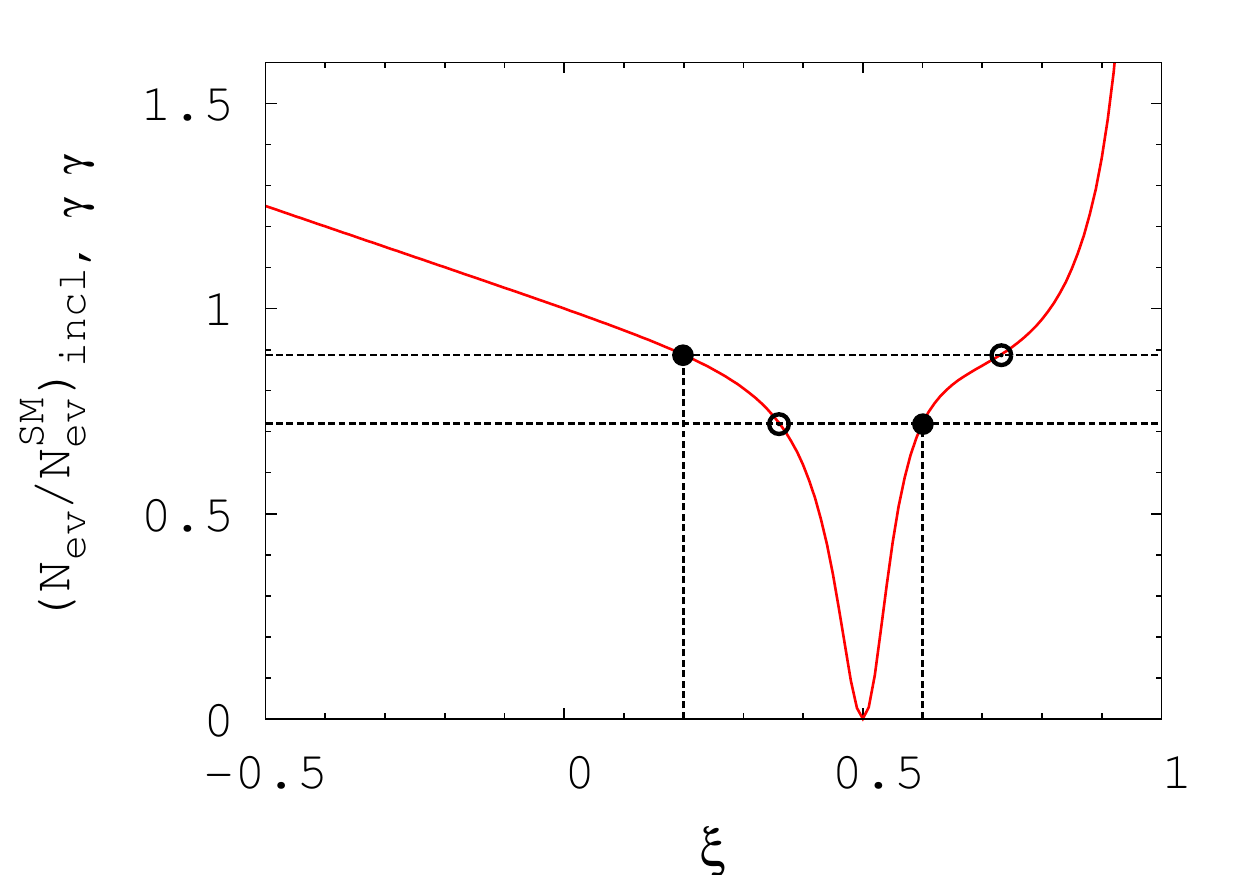}
\end{center}
\vspace*{-6mm}
\caption{The scaling functions for Higgs-strahlung $W H, H \to
         b\bar{b}$ and gluon fusion $gg \to H \to
         \gamma\gamma$. The full circle on the straight lines marks
         the true unique solution, while the open circles
         denote fake values which are different for the two channels.}
\label{fig:well}
\end{figure}

\begin{figure}[b]
\begin{center}
\includegraphics[width=0.40\textwidth]{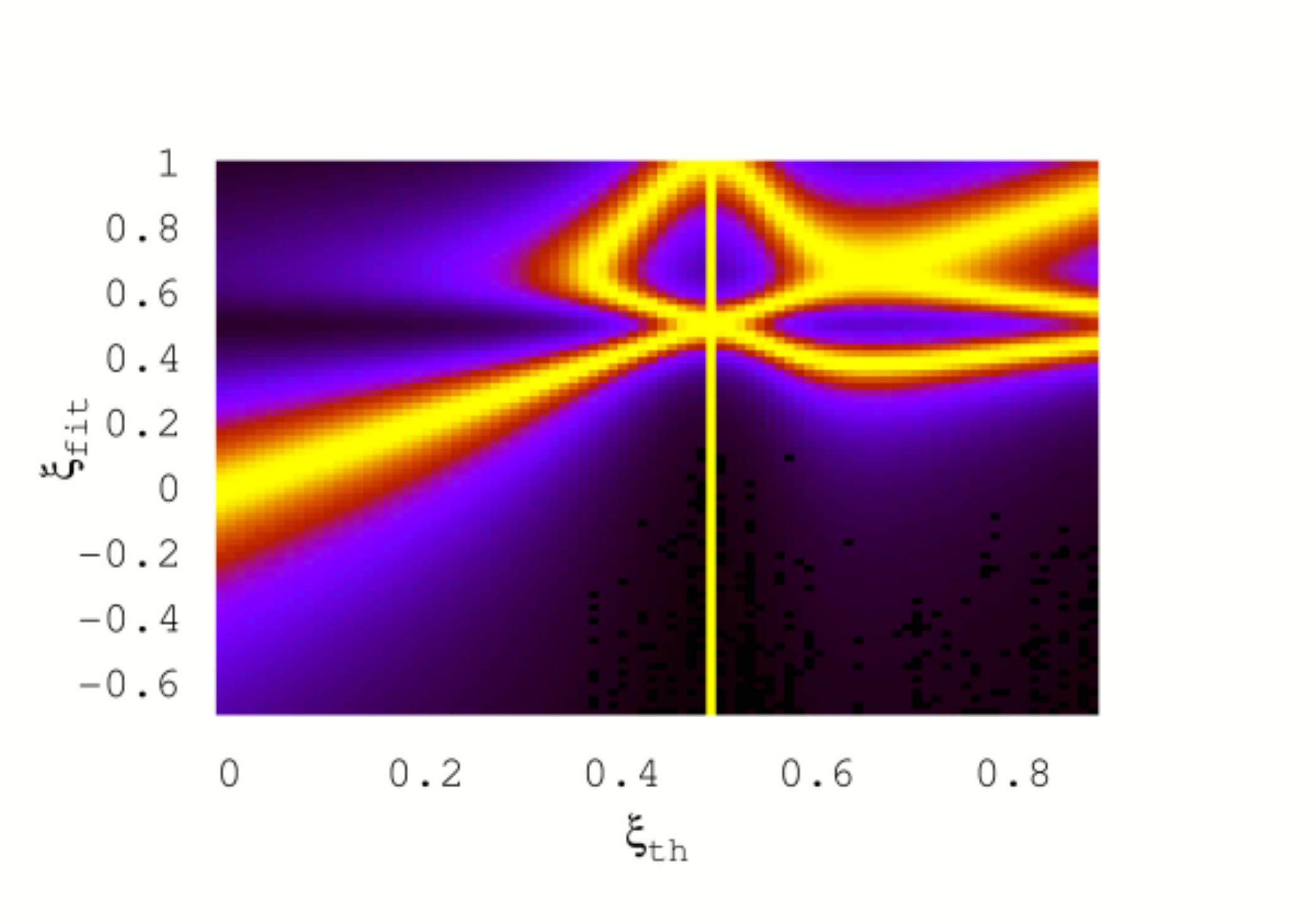}
\includegraphics[width=0.40\textwidth]{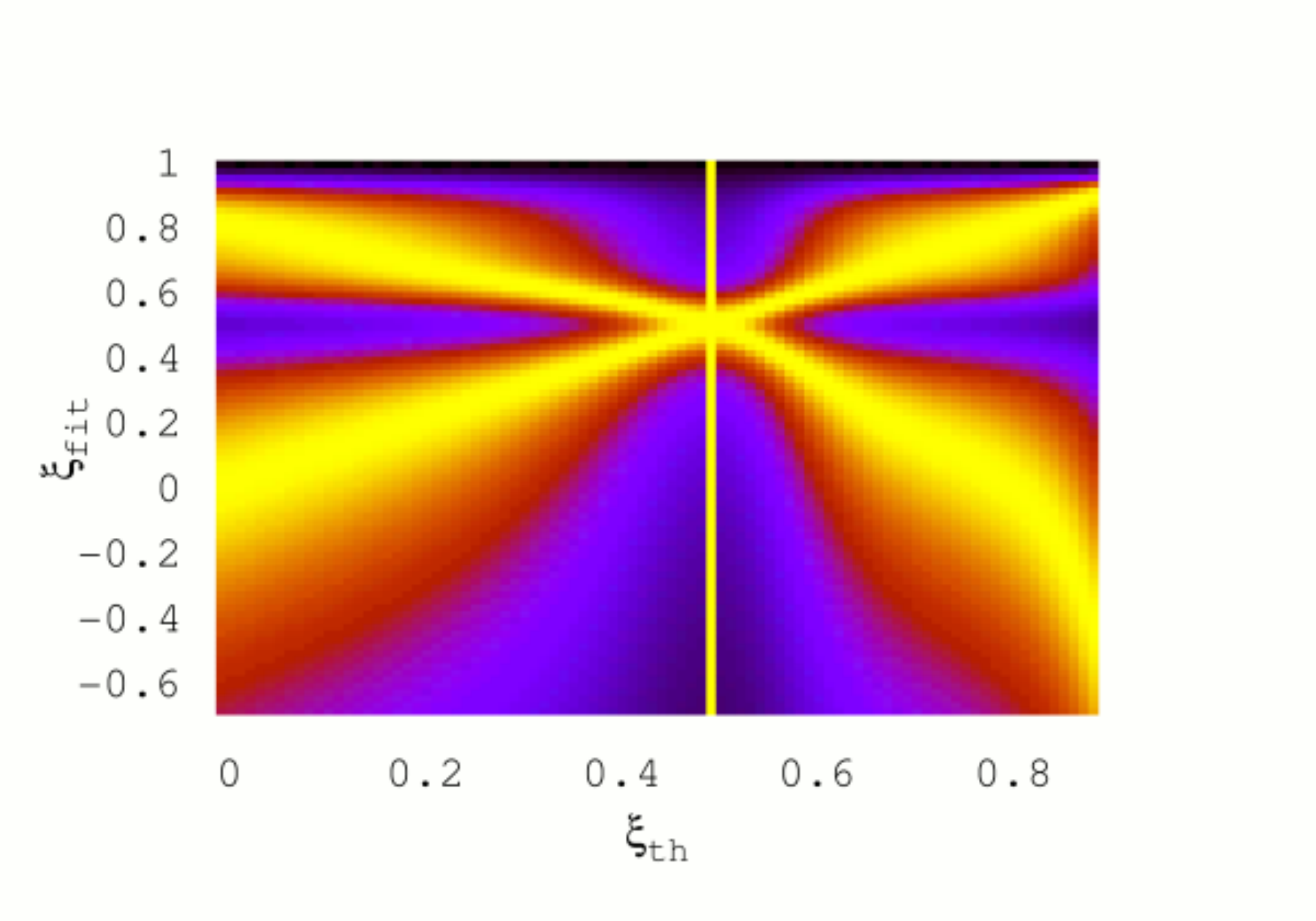} \\[-3mm]
\includegraphics[width=0.40\textwidth]{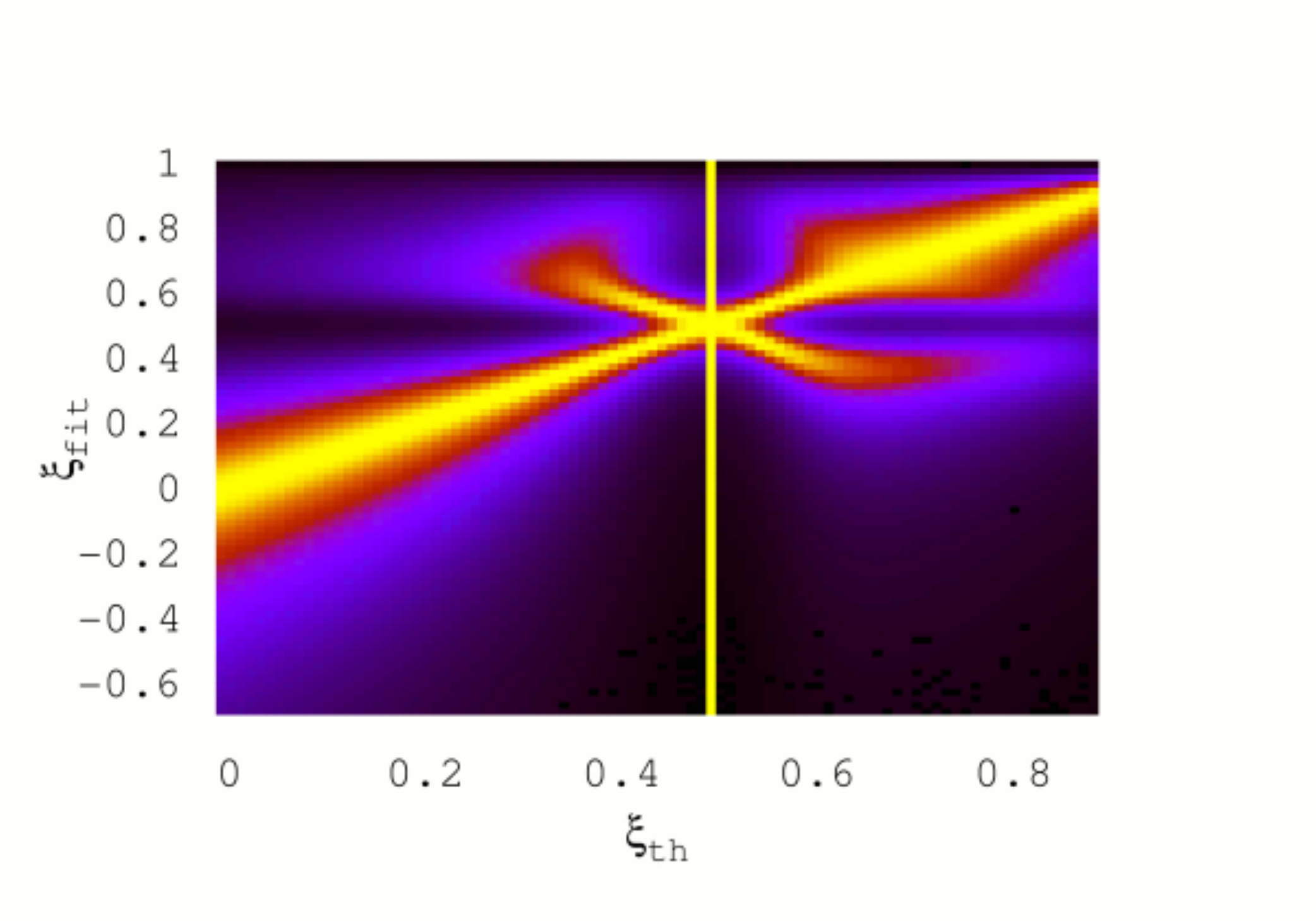}
\includegraphics[width=0.40\textwidth]{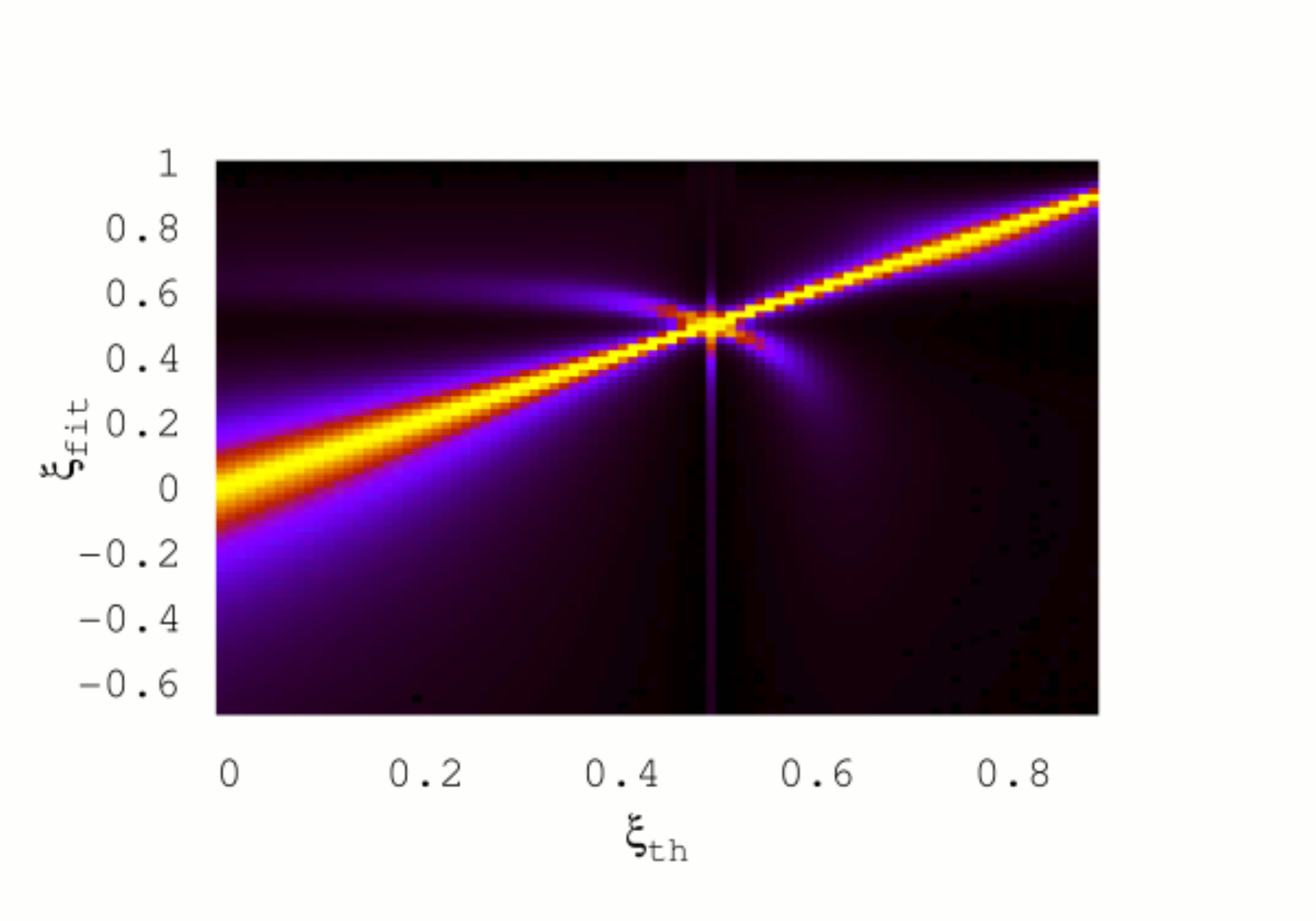}
\end{center}
\vspace*{-8mm}
\caption{Extracted values for $\xi$ as a function of the assumed
  $\xi_\text{th}$ for $M_H =$ 120 GeV and 30~fb$^{-1}$. We show 
  profile likelihoods, where we have neglected theory errors for clarity. 
  Upper: Higgs-strahlung plus
  $b\bar{b}$ decay [left], and gluon fusion plus $\gamma\gamma$ decay
  [right]. Lower: Combination of the two channels [left] and including all
  available channels [right].}
\label{fig:scaling}
\end{figure}

\section{Scaling of cross sections in a strongly interacting Higgs scenario}
\label{sec:appaltsol}

For the un-aligned shifts of Higgs-gauge and Yukawa couplings the
scaling of the cross sections with $\xi$ does not follow a simple
pattern.  However, the main features can be illustrated for two
characteristic leading channels, Higgs-strahlung $q \bar{q}' \to WH, H
\to b\bar{b}$, and gluon-fusion $gg \to H \to \gamma\gamma$. 
Ignoring resummation effects in this qualitative discussion, the
corresponding scaling functions, normalized to unity at the Standard
Model value $\xi = 0$, can be approximated as
\begin{alignat}{5}
\left( \frac{N_\text{ev}}{N_\text{ev}^\text{SM}} \right)_{WH, bb}
&= \frac{1}{\mathcal{N}_{WH,bb}} \,
   \frac{(1-\xi) \, \dfrac{(1-2\xi)^2}{(1-\xi)}}
              {\dfrac{(1-2\xi)^2}{(1-\xi)} + \gamma (1-\xi)} \; \notag \\[2mm]
\left( \frac{N_\text{ev}}{N_\text{ev}^\text{SM}} \right)_{gg, \gamma \gamma} 
&= \frac{1}{\mathcal{N}_{gg,\gamma\gamma}} \,
   \dfrac{\dfrac{(1-2\xi)^2}{(1-\xi)} \, 
                     \left[ \dfrac{(1-2\xi)^2}{(1-\xi)} + \beta_1 (1-2\xi)
                            + \beta_2 (1-\xi) \right] }
                    {\dfrac{(1-2\xi)^2}{(1-\xi)} + \gamma (1-\xi)} \; .
\end{alignat}
The parameter $\gamma \sim 0.05$ accounts for the admixture of $H \to
WW$ and $ZZ$ decays compared to $H \to b\bar{b}$ for Higgs masses
close to 120~GeV while $\beta_{1,2} \sim -8.9,\, 19.8$ parametrize
top- and $W$-loop contributions for Higgs decays to photons,
including their destructive interference.\medskip

As shown in Figure~\ref{fig:well} the scaling function for
Higgs-strahlung plus $b\bar{b}$ decays falls off straight, apart from
a narrow well close to $\xi = 1/2$. There, the rates for observing a
Higgs boson are driven to zero. The scaling function for gluon fusion
and $\gamma\gamma$ decay behaves similarly, except above the well,
where the scaling function diverges for $\xi \to 1$.  The
multiplicities of the solutions are illustrated for $\xi = 0.2$ and
$\xi = 0.6$ by the two dashed lines.  Depending on the value of $\xi$,
either one or up to three solutions correspond to given values of the
scaling functions. By combining the two sets of the solutions for the
two channels the individual ambiguities are resolved: the fake
solutions for $\xi$ [open circles] are eliminated and only the true
value [full circle] remains.\bigskip

This theoretical picture is realized only for large event
numbers. Otherwise, remnants of the fake solutions cannot be
eliminated completely. This is illustrated in Figure~{\ref{fig:scaling}}
where we present the log-likelihood bands for $\xi$. In the upper
panels we show the individual Higgs-strahlung and gluon-fusion
channels, including their individual fake solutions.  In the lower
panels we first combine the two channels, strongly reducing the fake
solutions. In the bottom right panel we include all available
channels, eliminating entirely the fake solutions. Only the width
around the true value in two preferred directions remains.


\end{document}

%% file: declare.tex
\def\tablename{Table}
\def\figurename{Figure}

\def\gtot{\Gamma_\text{tot}}
\def\brinv{\text{BR}_\text{inv}}
\def\brsm{\text{BR}^\text{SM}}
\def\bratio{\mathcal{B}_\text{inv}}
\def\as{\alpha_s}
\def\az{\alpha_0}
\def\gz{g_0}
\def\qlb{\overline{Q}_L}
\def\qrb{\overline{Q}_R}
\def\ql{Q_L}
\def\qr{Q_R}
\def\elb{\overline{L}_L}
\def\erb{\overline{L}_R}
\def\el{L_L}
\def\er{L_R}
\def\w{\vec{w}}
\def\sdag{\Sigma^{\dag}}
\def\s{\Sigma}
\newcommand{\psib}{\overline{\psi}}
\newcommand{\Psib}{\overline{\Psi}}
\newcommand\one{\leavevmode\hbox{\small1\normalsize\kern-.33em1}}
\newcommand{\Mpl}{M_\mathrm{Pl}}
\newcommand{\p}{\partial}
\newcommand{\lag}{\mathcal{L}}
\newcommand{\qqquad}{\qquad \qquad}
\newcommand{\qqqquad}{\qquad \qquad \qquad}

\newcommand{\qb}{\bar{q}}
\newcommand{\matx}{|\mathcal{M}|^2}
\newcommand{\really}{\stackrel{!}{=}}
\newcommand{\msbar}{\overline{\text{MS}}}
\newcommand{\qns}{f_q^\text{NS}}
\newcommand{\lqcd}{\Lambda_\text{QCD}}
\newcommand{\met}{\slashchar{E}_T}

\newcommand{\mev}{{\ensuremath\rm MeV}}
\newcommand{\gev}{{\ensuremath\rm GeV}}
\newcommand{\tev}{{\ensuremath\rm TeV}}
\newcommand{\fb}{{\ensuremath\rm fb}}
\newcommand{\ab}{{\ensuremath\rm ab}}
\newcommand{\pb}{{\ensuremath\rm pb}}
\newcommand{\sign}{{\ensuremath\rm sign}}
\newcommand{\ifb}{{\ensuremath\rm fb^{-1}}}

\def\slashchar#1{\setbox0=\hbox{$#1$}           
   \dimen0=\wd0                                 
   \setbox1=\hbox{/} \dimen1=\wd1               
   \ifdim\dimen0>\dimen1                        
      \rlap{\hbox to \dimen0{\hfil/\hfil}}      
      #1                                        
   \else                                        
      \rlap{\hbox to \dimen1{\hfil$#1$\hfil}}   
      /                                         
   \fi}
\newcommand{\dslash}{\slashchar{\partial}}
\newcommand{\Dslash}{\slashchar{D}}

\def\eg{{\sl e.g.} \,}
\def\ie{{\sl i.e.} \,}